\documentclass[twocolumn,dvipsnames,twocolappendix]{aastex631}
\usepackage{amsmath,amsfonts,amssymb,graphicx,chngcntr,multirow, float,booktabs}
\usepackage[]{hyperref}
\hypersetup{colorlinks=true}
\usepackage{threeparttablex}
\usepackage{wasysym}
\usepackage{arydshln}
\usepackage{mathrsfs}
\usepackage{amssymb}

\begin{document}

\title{SRG/eROSITA No. 5: \\ Discovery of quasi-periodic eruptions every $\sim3.7$\,days from a galaxy at $z>0.1$}

\author[0000-0003-4054-7978]{R. Arcodia}\thanks{NASA Einstein Fellow}
\affiliation{Kavli Institute for Astrophysics and Space Research, Massachusetts Institute of Technology, Cambridge, MA 02139, USA}

\author{P. Baldini}
\affiliation{Max-Planck-Institut für extraterrestrische Physik (MPE), Gießenbachstraße 1, 85748 Garching bei München, Germany}

\author{A. Merloni}
\affiliation{Max-Planck-Institut für extraterrestrische Physik (MPE), Gießenbachstraße 1, 85748 Garching bei München, Germany}

\author{A. Rau}
\affiliation{Max-Planck-Institut für extraterrestrische Physik (MPE), Gießenbachstraße 1, 85748 Garching bei München, Germany}

\author{K. Nandra}
\affiliation{Max-Planck-Institut für extraterrestrische Physik (MPE), Gießenbachstraße 1, 85748 Garching bei München, Germany}

\author{J. Chakraborty}
\affiliation{Kavli Institute for Astrophysics and Space Research, Massachusetts Institute of Technology, Cambridge, MA 02139, USA}

\author{A. J. Goodwin}
\affiliation{International Centre for Radio Astronomy Research – Curtin University, GPO Box U1987, Perth, WA 6845, Australia}

\author{M. J. Page}
\affiliation{Mullard Space Science Laboratory, University College London, Holmbury St Mary, Dorking, Surrey, RH5 6NT, UK}

\author{J. Buchner}
\affiliation{Max-Planck-Institut für extraterrestrische Physik (MPE), Gießenbachstraße 1, 85748 Garching bei München, Germany}

\author{M. Masterson}
\affiliation{Kavli Institute for Astrophysics and Space Research, Massachusetts Institute of Technology, Cambridge, MA 02139, USA}

\author{I. Monageng}
\affiliation{South African Astronomical Observatory, P.O. Box 9, Observatory 7935, Cape Town, South Africa}

\author{Z. Arzoumanian}
\affiliation{Astrophysics Science Division, NASA Goddard Space Flight Center, 8800 Greenbelt Road, Greenbelt, MD 20771, USA}

\author{D. Buckley}
\affiliation{South African Astronomical Observatory, P.O. Box 9, Observatory 7935, Cape Town, South Africa}

\author{E. Kara}
\affiliation{Kavli Institute for Astrophysics and Space Research, Massachusetts Institute of Technology, Cambridge, MA 02139, USA}

\author{G. Ponti}
\affiliation{INAF-Osservatorio Astronomico di Brera, Via E. Bianchi 46, I-23807 Merate (LC), Italy}
\affiliation{Max-Planck-Institut für extraterrestrische Physik (MPE), Gießenbachstraße 1, 85748 Garching bei München, Germany}
\affiliation{Como Lake Center for Astrophysics (CLAP), DiSAT, Università degli Studi dell’Insubria, via Valleggio 11, I-22100 Como, Italy}

\author{M. E. Ramos-Ceja}
\affiliation{Max-Planck-Institut für extraterrestrische Physik (MPE), Gießenbachstraße 1, 85748 Garching bei München, Germany}

\author{M. Salvato}
\affiliation{Max-Planck-Institut für extraterrestrische Physik (MPE), Gießenbachstraße 1, 85748 Garching bei München, Germany}

\author{K. Gendreau}
\affiliation{Astrophysics Science Division, NASA Goddard Space Flight Center, 8800 Greenbelt Road, Greenbelt, MD 20771, USA}

\author{I. Grotova}
\affiliation{Max-Planck-Institut für extraterrestrische Physik (MPE), Gießenbachstraße 1, 85748 Garching bei München, Germany}

\author{M. Krumpe}
\affiliation{Leibniz-Institut für Astrophysik Potsdam (AIP), An der Sternwarte 16, 14482 Potsdam, Germany}



\begin{abstract}

Quasi-periodic eruptions (QPEs) are repeating soft X-ray bursts from the nuclei of galaxies, tantalizingly proposed to be extreme mass ratio inspirals. Here, we report the discovery of a new galaxy showing X-ray QPEs, the fifth found through a dedicated blind search in the \emph{SRG}/eROSITA all-sky survey data, hereafter named eRO-QPE5. Its QPE duration ($t_{\rm dur}\sim0.6$\,d), recurrence time ($t_{\rm recur}\sim3.7\,$d), integrated energy per eruption ($\sim3.4 \times 10^{47}\,$erg), and black hole mass ($M_{\rm BH}=2.9^{+5.4}_{-2.2}\times10^7\,M_{\astrosun}$) sit at the high end of the known population. Like other eROSITA or X-ray-discovered QPEs, no previous or concurrent optical-IR transient is found in archival photometric datasets, and the optical spectrum looks almost featureless. With a spectroscopic redshift of $0.1155$, eRO-QPE5 is the most distant QPE source discovered to date. Given the number of recent discoveries, we test for possible correlations and confirm a connection between $t_{\rm dur}$ and $t_{\rm recur}$, while we do not find any significant correlation involving either $M_{\rm BH}$ or the QPE temperature. The slope of the $t_{\rm dur}-t_{\rm recur}$ relation ($1.14\pm0.16$) is roughly consistent with predictions from star-disk collision models, with a preference for those that suggest that QPEs are powered by stellar debris streams around the orbiter. Considering this and previous discoveries, eROSITA has proved extremely successful in finding many QPE candidates given its grasp, namely its sensitivity and large field of view, and scanning capabilities over the full sky. We advocate the need of sensitive wide-area and time-domain oriented surveys from future-generation soft X-ray missions. 
\end{abstract}



\defcitealias{Arcodia+2021:eroqpes}{A21}
\defcitealias{Arcodia+2024:eroqpes}{A24}

\section{Introduction}
\label{sec:intro}

Quasi-periodic eruptions (QPEs) are soft X-ray flares that repeat on the timescales of a few hours to a few days \citep{Miniutti+2019:qpe1,Giustini+2020:qpe2,Arcodia+2021:eroqpes,Arcodia+2024:eroqpes,Chakraborty+2021:qpe5cand,Chakraborty+2025:upj,Quintin+2023:tormund,Bykov+2024:tormund,Nicholl+2024:qiz,Hernandez-Garcia+2025:ansky}, originating from massive black holes (MBHs) with masses $M_{\rm BH}\approx10^{5-7.5}$ \citep[e.g.,][]{Wevers+2022:host,Wevers+2024:host} in galactic nuclei. When the galactic nuclei of QPE sources are not flaring, they show a quiescent optically thick thermal soft X-ray spectrum (with peak effective temperature $kT \sim30-80\,$eV) that was promptly interpreted as arising from the inner accretion disk of the nuclear MBH. This is supported by the recent detection of a nuclear UV point-like source in eRO-QPE2, AT2019qiz, and GSN\,069 \citep{Nicholl+2024:qiz,Wevers+2025:sed,Guolo+2025:sed,Guolo+2025:gsnlongterm}. The most tested theory\footnote{For alternative recent models, see for instance \citet{Kaur+2023:Bdisks,Pan+2023:instab,Middleton+2025:prec,D'Orazio+2025:cbdtde}.} reproducing many observed QPE properties (although not all) involves an extreme mass-ratio inspiral (EMRI), and suggests that QPEs are triggered by a stellar-mass object in a low-eccentricity orbit repeatedly colliding, typically once or twice per orbit, with the accretion flow of the primary MBH \citep{Xian+2021:collisions,Linial+2023:qpemodel,Franchini+2023:qpemodel,Tagawa+2023:qpemodel,Zhou+2024:qpemodel,Vurm+2024:emission,Huang+2025:sims}. This framework requires the smaller orbiter to slowly evolve primarily through gravitational wave (GW) emission down to the inferred current orbital configuration \citep[e.g.,][]{Linial+2022:segreg,Rom+2025:segregation}, and that the accretion flow with which it interacts is newborn, perhaps fed by a tidal disruption event (TDE) of an independent object \citep{Linial+2023:qpemodel,Franchini+2023:qpemodel}. Indeed, observations suggest that the accretion flow in QPE sources is compact \citep[e.g.,][]{Miniutti+2023:gsnrebr,Guolo+2025:sed,Wevers+2025:sed} and of recent origin \citep[e.g.,][]{Patra+2024:hst}, and we now know of at least three QPE sources associated with previous spectroscopically-confirmed optically-selected TDEs \citep{Quintin+2023:tormund,Bykov+2024:tormund,Nicholl+2024:qiz,Chakraborty+2025:upj}, in addition to more circumstantial evidence connecting QPEs and TDEs \citep{Chakraborty+2021:qpe5cand,Sheng+2021:tdeconn,Miniutti+2023:gsnrebr,Arcodia+2024:eroqpes,Wevers+2024:host,Gilbert+2024:hosts,Kosec+2025:specs} and recent application of the relativistic time-dependent accretion model \texttt{FitTeD} \citep{Mummery+2024:fitted} to GSN\,069 \citep{Guolo+2025:gsnlongterm}. Furthermore, one of the most recent additions to the QPE population is associated with a previous accretion event ``awakening'' the MBH, but which has not been unambiguously associated with a TDE \citep{SanchezSaez+2024:ansky,Hernandez-Garcia+2025:ansky}. This might suggest that QPEs require a recent accretion event of whatever origin, for which TDEs provide a natural channel. We note that some very recent works have casted some doubt on the first collision prescriptions \citep[e.g.,][]{Guo+2025:coll,Guolo+2025:gsnlongterm,Mummery+2025:collisions,Yao+2025:simul,Linial+2025:streams} and to date no single QPE EMRI model is able to reproduce all QPE sources.

Notably, EMRIs are expected to emit GWs and, depending on the exact nature of the orbiter, will be detectable by the recently-adopted LISA \citep{Amaro-Seoane+2007:emris} and TianQin \citep{Luo2016:tianqin} missions, or by possible future-generation $\mu$-Hz detectors \citep{Sesana+2021:muAres}. Confirming this interpretation and constraining the type and mass of the orbiter triggering QPEs may thus have significant consequences for future multi-messenger synergies with X-ray missions, such as AXIS \citep{Reynolds+2023:axis} and NewAthena \citep{Nandra+2013:athena,Cruise+2025:newathena}. For this, some first tests and consistency checks using existing QPE timing data \citep{Chakraborty+2024:ero1,Arcodia+2024:ero2timing,Pasham+2024:qpe1,Pasham+2024:ero2,Miniutti+2025:gsn,Zhou+2024:longterm,Guo+2025:coll,Mummery+2025:collisions,Xian+2025:coll}, as well as disk size \citep{Nicholl+2024:qiz,Wevers+2025:sed,Chakraborty+2025:upj,Guolo+2025:sed,Guolo+2025:gsnlongterm}, have been performed. However, at this stage it is also important to find more QPE sources to sample the diversity of the population, so far limited in number statistics. This goal is hampered by the fact that to date QPE flares have only been revealed in the soft X-rays, and blind systematic searches have only been possible with \emph{SRG/eROSITA} \citep{Arcodia+2021:eroqpes,Arcodia+2024:eroqpes}, which halted its all-sky survey in early 2022.

Here, we report the discovery of a new galaxy which showed significant high-amplitude variability in archival 2019-2021 \emph{SRG}/eROSITA data, later confirmed to show X-ray quasi-periodic eruptions by \emph{Swift}/XRT, \emph{NICER} and \emph{XMM-Newton}. The X-ray source eRASSt J032543.2-451244 (or J0325 in short) is the fifth QPE discovery obtained through a blind search in \emph{SRG}/eROSITA all-sky survey data, and it is hereafter named eRO-QPE5. 

\section{X-ray analysis and results}
\label{sec:xray}

\begin{figure*}[t]
		\centering
        \includegraphics[width=0.99\textwidth]{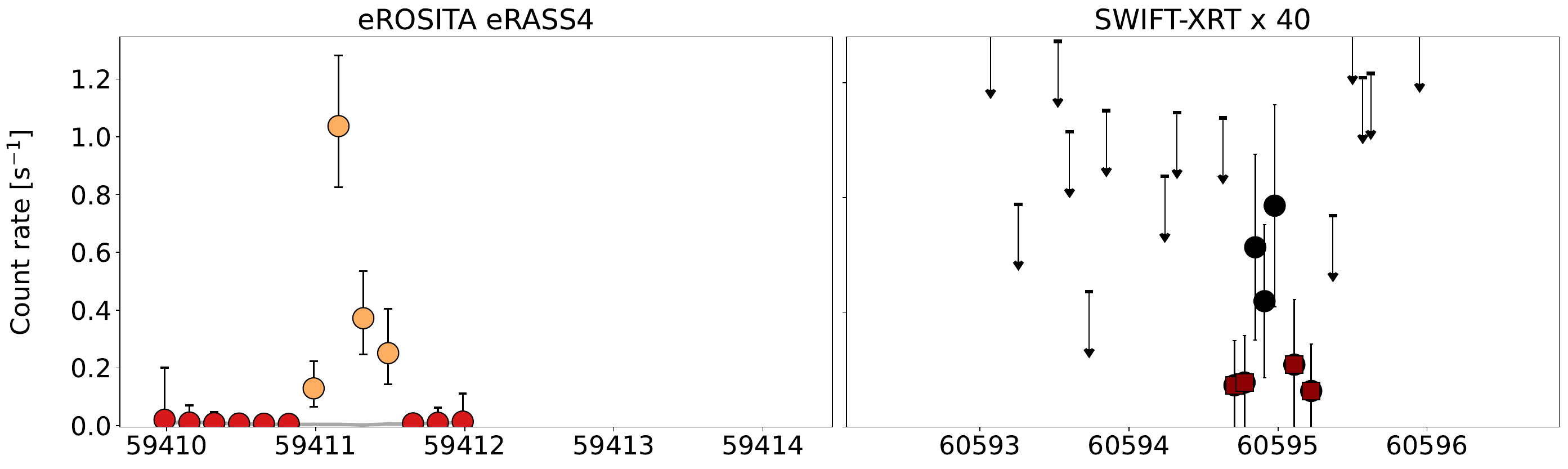}
        \includegraphics[width=0.99\textwidth]{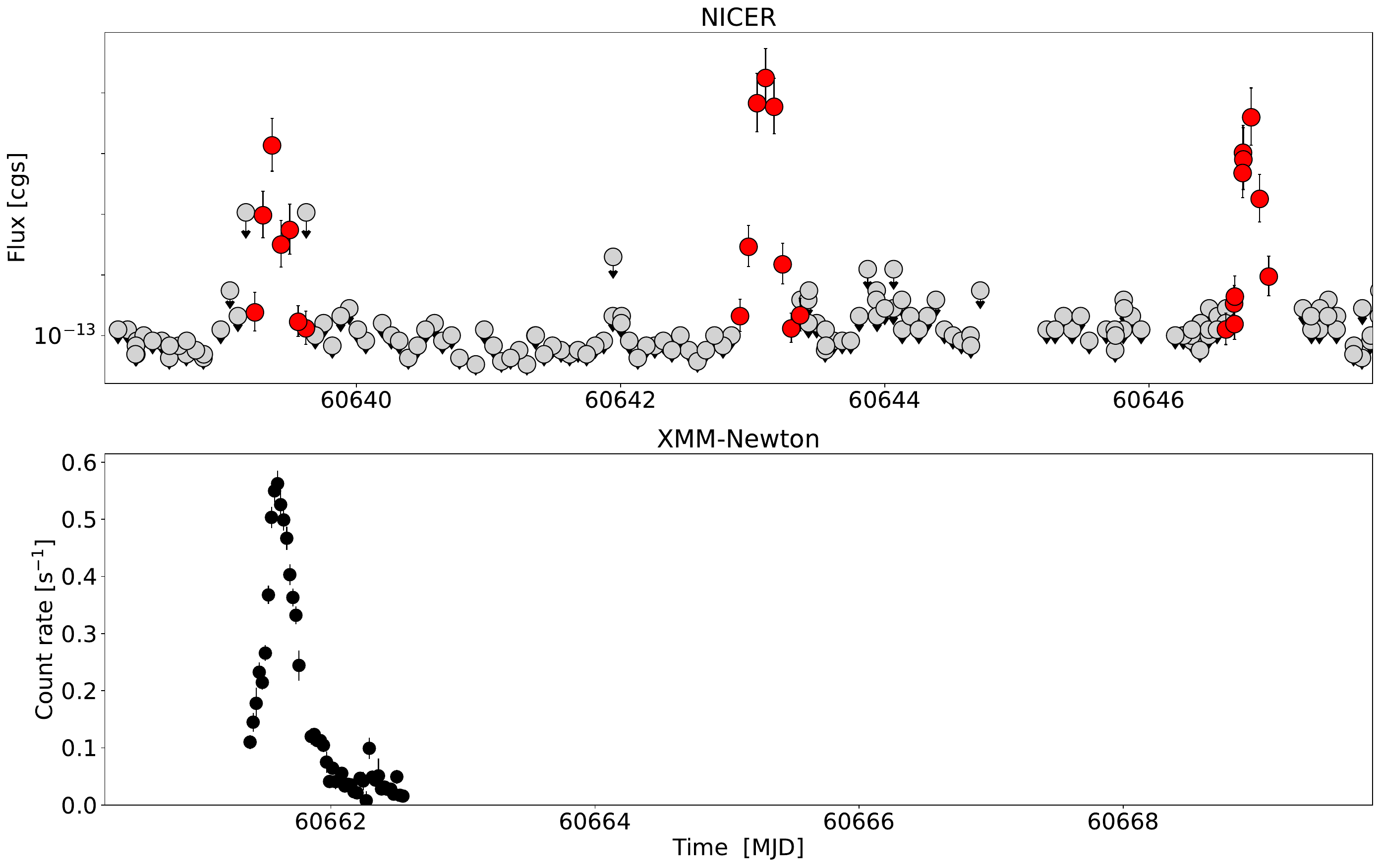}
    	\caption{X-ray light curve of eRO-QPE5 at different epochs, from top to bottom: eROSITA eRASS4 $0.2-2.3\,$keV light curve (detected eROdays in orange; top left panel) and \emph{Swift}/XRT $0.3-2.0\,$keV light curve (marginal detections in dark red; top right panel); \emph{NICER} $0.4-2.0\,$keV light curve (detections in red); \emph{XMM-Newton} EPICpn $0.2-2.0\,$keV light curve.}
		\label{fig:lc}
\end{figure*}

Here, we present a summary of our analysis of the multi-mission X-ray data on J0325 (see Appendix~\ref{sec:xray_proc} for more details). We have presented in previous work (\citetalias{Arcodia+2024:eroqpes}; \citealp{Arcodia+2024:rates}) our algorithm to look for repeated high-amplitude variability in the all-sky survey data of eROSITA \citep{Predehl+2021:erosita}, which is the primary instrument on the Spectrum-Roentgen-Gamma (SRG, \citealp{Sunyaev+2021:srg}) mission. eROSITA light curves were rebinned to eROdays (i.e. one data point per eROSITA snapshot) using the eRebin tool\footnote{\href{https://github.com/rarcodia/eRebin}{https://github.com/rarcodia/eRebin}} (\citetalias{Arcodia+2021:eroqpes,Arcodia+2024:eroqpes}; \citealp{Buchner+2022A&A...661A..18B}). J0325 triggered our QPE search algorithm in the last of the four completed all-sky surveys (eRASS4), as it was detected only in four consecutive visits (i.e., eROdays) that appeared in a flare-shaped profile with a fast rise and slow decay (Fig.~\ref{fig:lc}, top left). The detected phase (orange points in the top left panel of Fig.~\ref{fig:lc}) appeared soft ($\Gamma > 3.3$) when fitted by a simple power law, and fitting with a thermal \texttt{zbbody} model yielded $kT=111^{+14}_{-11}\,$eV and a flux, corrected for Galactic absorption, of $F_{0.2-2.0\,\rm keV} = 1.3^{+0.3}_{-0.3}\times 10^{-12}\,$erg\,s$^{-1}$\,cm$^{-2}$. The non detections in eRASS4 (red points in the top left panel of Fig.~\ref{fig:lc}) result in a $3\sigma$ upper limit at $F_{0.2-2.0\,\rm keV} < 6.7 \times 10^{-14}\,$erg\,s$^{-1}$\,cm$^{-2}$ (\texttt{diskbb} model). J0325 was marginally detected in all of the previous three eRASS surveys, with each also showing tentative evidence for variability (see Sect.~\ref{sec:ero_proc}). The association of J0325 with a galaxy (Sect.~\ref{sec:opt}) and the promising X-ray variability and spectral shape prompted further follow-up observation to confirm a QPE origin. 

The source was observed by both \emph{Swift}/XRT and \emph{NICER} in guest observer time in the second half of 2024. \emph{Swift}/XRT took data during Cycle 20 with program 2023085 (PI: Arcodia) for a total of $\sim13.9$\,ks in a baseline of $\sim2.9$\,days. Single snapshots with exposures varying between $\approx300-400$\,s were taken with a typical cadence of $\sim1.6$\,h. We extracted \emph{Swift}/XRT products using the \texttt{swifttools.ukssdc} python module\footnote{\href{https://www.swift.ac.uk/API}{https://www.swift.ac.uk/API}}, such as the $0.3-2.0\,$keV light curve binned per snapshots (top right panel in Fig.~\ref{fig:lc}). We performed a barycenter correction with the task \texttt{barycorr}. As shown in the top-right panel of Fig.~\ref{fig:lc}, \emph{Swift}/XRT detected a single flare with similar characteristics and apparent duration as the eRASS4 flare. We verified the detections around MJD$\sim60595$ performing aperture photometry on the image extracted between MJDs $60594.7-60595.2$, obtaining a no-source binomial probability (see \citealp{Arcodia+2024:ap_photom} for a definition and method description) of $P_b\sim5\times10^{-11}$ that this signal is a background fluctuation\footnote{While this value is not calibrated against \emph{Swift}/XRT simulations, a significant detection threshold is achieved at around $P_b\approx10^{-4}$ for typical X-ray imagers \citep{Luo+2017:chandra,Arcodia+2024:ap_photom}.}. We converted count rates to flux using \texttt{WebPIMMS}: for the QPE phase (MJDs $\sim 60594.8-60595.0$) we adopted a $kT\sim0.1\,$keV black body and obtained $F_{0.2-2.0\,\rm keV} = 8.0^{+1.6}_{-4.0}\times 10^{-13}\,$erg\,s$^{-1}$\,cm$^{-2}$; for the quiescence phase (e.g. MJD$\sim60593.87$) we adopted a $kT\sim0.05\,$keV black body and obtained a $1\sigma$ upper limit at $F_{0.2-2.0\,\rm keV} < 2.3 \times 10^{-13}\,$erg\,s$^{-1}$\,cm$^{-2}$.

\emph{NICER} data were first taken between October 16 and 25 2024 (OBSIDs 7608020101-7608020102 and 7608020201-7608021001), but mostly during orbit day, which allows detection at shallower fluxes than orbit night due to increased low-energy contamination. A new observation set obtained as target of opportunity between November 24 and December 5 2024 (OBSIDs 7608020103-7608020114), with data taken mostly during \emph{NICER} orbit night, detected three consecutive eruptions separated by $\approx3.7$\,d (Fig.~\ref{fig:lc}, medium panel). We triggered \emph{XMM-Newton} in discretionary time (ObsID 0954190401, starting on 17 December 2024) to detect an eruption with high signal-to-noise ratio (Fig.~\ref{fig:lc}, bottom panel). We fitted the better-sampled \emph{NICER} and \emph{XMM-Newton} light curves to infer the average duration ($t_{\rm dur}$) and recurrence time ($t_{\rm recur}$), adopting a flare model composed by a double exponential \citep{Arcodia+2022:ero1_timing} for the fit (see Appendix~\ref{sec:xray_proc} for more details). The mean and standard deviation of the fitted rise-to-decay durations (defined from the times at which the fitted profile reaches $1/e^3$ of the peak) of the three \emph{NICER} bursts are $0.69\pm0.09$\,d, $0.65\pm0.07$\,d, and $0.50\pm0.05$\,d, respectively. While the third fitted duration is shorter than the first two, this burst has a gap with no data for $\sim0.25\,$d at the end of the burst. Thus, the decay and consequently the duration is not sampled as well as the first two eruptions and this might not be an intrinsic scatter in the burst duration. The inferred rise-to-decay duration of the \emph{XMM-Newton} is $0.73\pm 0.02\,$d, thus somewhat larger than those inferred by \emph{NICER}. This might hint at biases when instruments with different sensitivities are used to fit the early rise and late decays of eruptions. To test this, we also inferred the near-peak duration (between times at which the intensity reaches $1/e$ of the peak value, instead of $1/e^3$). The values of the three \emph{NICER} and one \emph{XMM-Newton} bursts are $0.27\pm0.04$\,d, $0.25\pm0.03$\,d, $0.20\pm0.02$\,d, and $0.29\pm0.01$\,d, respectively, thus indeed more similar. In summary, the average rise-to-decay duration (defined from the times at which the fitted profile reaches $1/e^3$ of the peak) across these four observed bursts is $0.64\pm0.11\,$d, while the average near-peak duration (from the times at which the fitted profile reaches $1/e$ of the peak) is instead $0.25\pm0.04\,$d. These two duration estimates result in a duty cycle of $\sim 17\%$ and $\sim 7\%$, respectively. We computed the total integrated energy of the burst from the rise-to-decay profiles, which is $(3.0\pm0.3) \times 10^{47}\,$erg, $(4.3\pm0.3) \times 10^{47}\,$erg, and $(2.8\pm0.2) \times 10^{47}\,$erg, for the three \emph{NICER} eruptions, respectively, with an average value of $(3.4\pm0.7) \times 10^{47}\,$erg. $t_{\rm recur}$ can only be computed with \emph{NICER} data, and we estimate a mean and standard deviation of the two recurrence times at $3.72\pm0.01$\,d, $3.69\pm0.01$\,d, respectively. The average recurrence for eRO-QPE5 is thus $3.70\,$d with a standard deviation of $0.02\,$d. This scatter of $\sim0.5\%$ on the burst arrival time is remarkably low, the lowest among the known QPE sources which typically show significant variations (\citetalias{Arcodia+2021:eroqpes,Arcodia+2024:eroqpes}; \citealp{Miniutti+2019:qpe1,Giustini+2020:qpe2,Nicholl+2024:qiz}).

\begin{figure}[tb]
		\centering
		\includegraphics[width=0.99\columnwidth]{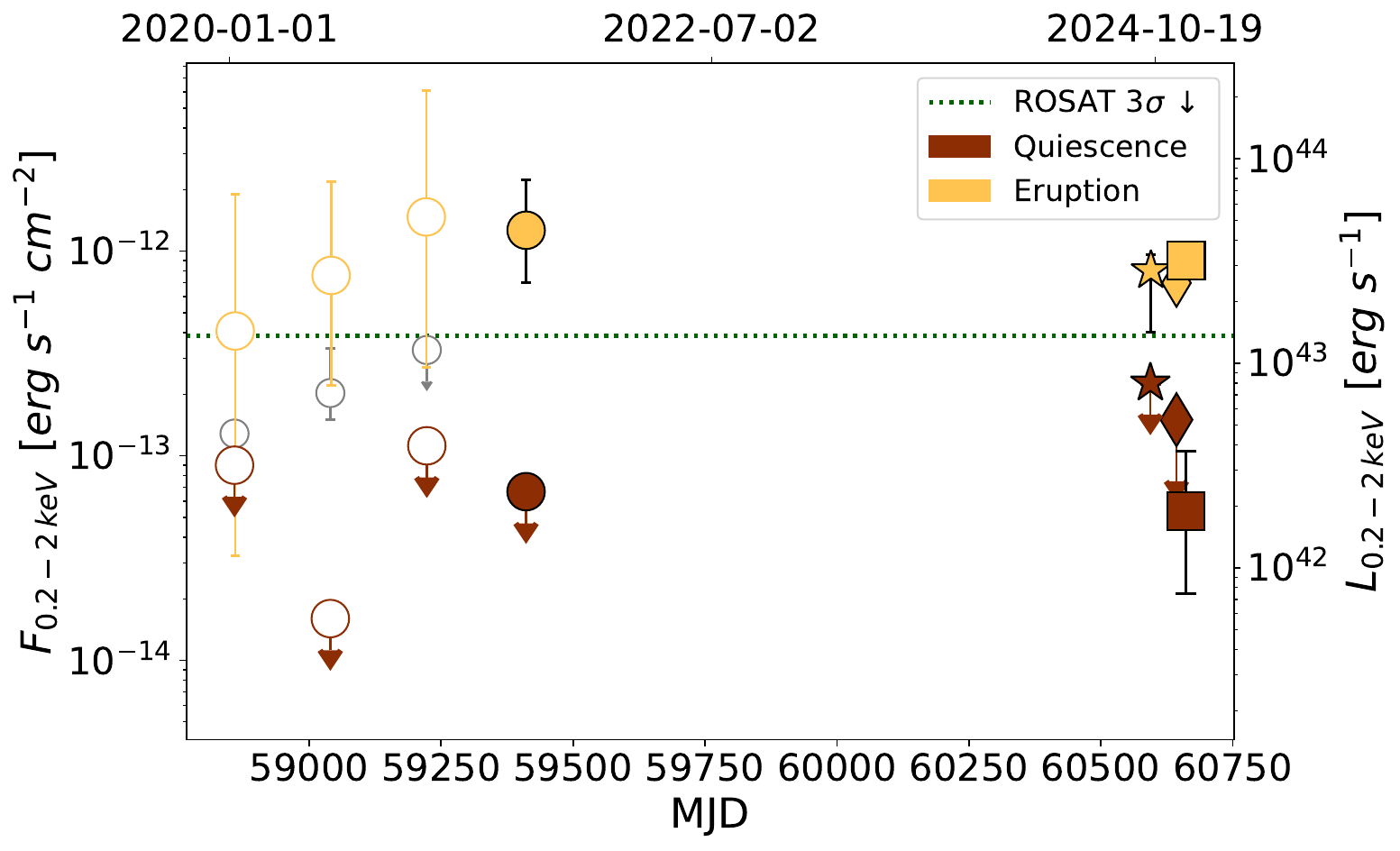}
		\caption{Long-term evolution of J0325, highlighting flux states both in quiescence and in the bright state (which includes the former). eROSITA data points are shown as circles, squares for \emph{XMM-Netwon}, diamonds for \emph{NICER} and stars for \emph{Swift}/XRT. eROSITA epochs with uncertain phase identification are shown with empty symbols, and the flux of the full exposure is shown in gray (see more details in Sect.~\ref{sec:ero_proc}). All uncertainties shown are $3\sigma$. The dotted gray horizontal line highlights a \emph{ROSAT} archival upper limit.}
		\label{fig:longterm}
\end{figure}

Finally, we show the long-term flux evolution of J0325 inferred with eROSITA, \emph{XMM-Newton}, \emph{NICER} and \emph{Swift}/XRT in Fig.~\ref{fig:longterm}, separating flux states in quiescence and in eruption as indicated in the legend (see Sect.~\ref{sec:xray_proc} for more details). The X-ray quiescence was only detected in the deeper \emph{XMM-Newton} observation. We extracted it selecting good time intervals after the source decayed to $1/e^3$ of the peak value. Since this emission component is interpreted as the inner regions of a radiatively efficient accretion disk, we modeled it with a simple \texttt{diskbb} given the relatively low signal-to-noise ratio (but see a more accurate treatment below). In fact, the net exposure is only $\sim16.4$\,ks for the quiescence phase. We obtained $kT_{\rm disk}=kT_{\rm in}= 37^{+11}_{-7}\,$eV and an unabsorbed flux of $F_{0.2-2.0\,\rm keV} = 5.4^{+2.3}_{-1.8}\times 10^{-14}\,$erg\,s$^{-1}$\,cm$^{-2}$. QPE peak fluxes are instead estimated with an additive thermal component (see Table.~\ref{tab:spec}), and are overall compatible within uncertainties across multiple epochs possibly suggesting a relatively low dispersion in peak flux of the eruptions. Compared to eROSITA $\sim3.4$\,y earlier, the median value of the peak $F_{0.2-2.0\,\rm keV}$ is $\approx 30\%$ lower, although still compatible within 3$\sigma$.

\begin{figure}[tb]
		\centering
		\includegraphics[width=0.99\columnwidth]{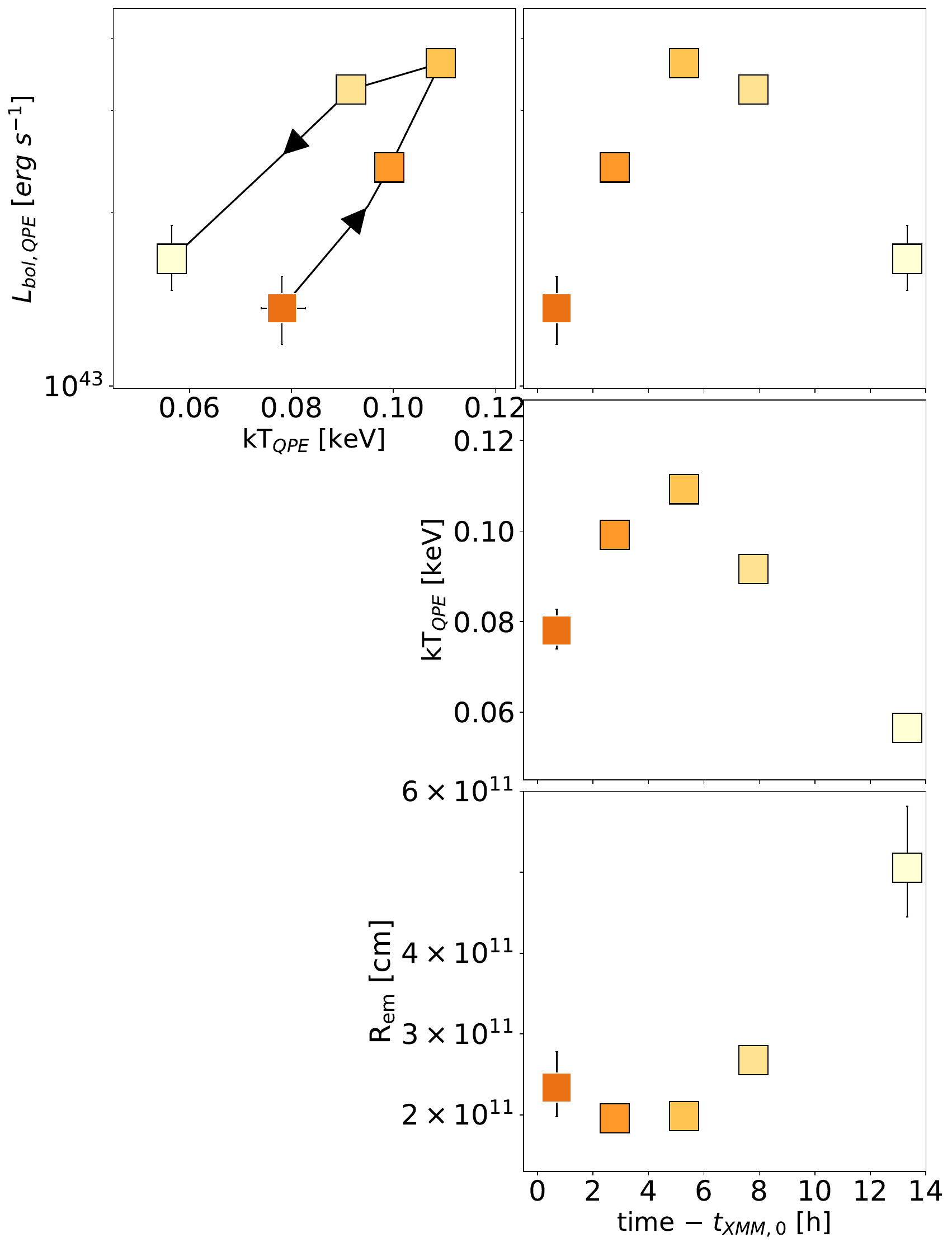}
		\caption{Spectral evolution of QPEs in eRO-QPE5 using XMM-Newton data. Time increases from darker to lighter colors. The top right (middle right) panel shows the QPE bolometric luminosity (temperature) evolution, while the top left panel shows their joint evolution. The bottom right panel shows the evolution of the emitting radius, assuming a thermal spectrum and spherical geometry. We note that the first rise phase (the darkest data point) is not fully captured by \emph{XMM-Newton}, thus it has to be interpreted with caution.}
		\label{fig:energy_evolution}
\end{figure}

\subsection{Spectral evolution during the eruptions}
\label{sec:energydep}

The sources eRO-QPE1-4 (\citealp{Arcodia+2022:ero1_timing,Arcodia+2024:ero2timing}; \citetalias{Arcodia+2024:eroqpes}), GSN\,069 \citep{Miniutti+2023:gsnrebr}, RX J1301 \citep{Giustini+2024:rxj}, XMM J0249 (\citealp{Chakraborty+2021:qpe5cand}; J. Chakraborty, priv. comm., for the energy behavior), AT2019qiz \citep{Nicholl+2024:qiz}, ZTF19acnskyy \citep{Hernandez-Garcia+2025:ansky}, and AT2022upj \citep{Chakraborty+2025:upj} all show a harder spectrum during the rise of the eruptions than during the decay, at a compatible count rate level\footnote{The candidate AT2019vcb does not have enough data to determine a recurrence time (currently bracketed within $8\,$h$-13\,$d) nor the QPE-like energy behavior for a full eruption \citep{Quintin+2023:tormund,Bykov+2024:tormund}, similarly to the ``rapid'' flares \citep{Pasham+2024:0230} in the X-ray repeating nuclear transient Swift J0230 \citep{Evans+2023:swift}, for which the main flares do not show the QPE-like energy behavior \citep{Guolo+2023:swift}.}. If the eruption spectra are fitted with a thermal model, this luminosity-temperature hysteresis implies that an expansion of the QPE emitting region occurs when going from the start to the end of the eruptions \citep{Miniutti+2023:gsnrebr,Chakraborty+2024:ero1}. Thus, this behavior may be the possible signature of a common emission mechanism that identifies bona fide QPEs among a growing number of diverse repeating nuclear transients. It is also qualitatively consistent with an expanding gas bubble being ejected after the putative collisions between the orbiter and the accretion disk \citep{Franchini+2023:qpemodel,Linial+2023:qpemodel}, and the first quantitative tests have recently been made with radiation transport calculations and spectral fitting (\citealp{Vurm+2024:emission,Chakraborty+2025:spec}; but see \citealp{Mummery+2025:collisions}). Here, we find that this characteristic spectral evolution is present in J0325 as well. We extracted spectra at five different flare phases (two during the rise, the peak, and two during the decay), separated by the times at which the source was at $\sim75\%$ and $\sim45\%$ of the peak. This choice is arbitrary, but the results are not significantly affected by different values. We show the spectral evolution, namely temperature, luminosity, and size of the emitting region, in Fig.~\ref{fig:energy_evolution}, together with the $L_{\rm bol,QPE}$--$kT_{\rm QPE}$ hysteresis plot. Details on the fitted values are shown in Table~\ref{tab:spec}. We note that the \emph{XMM-Newton} light curve starts during the rise of the eruption, so that the first rise phase is incomplete. Thus, we highlight the first bin with a different edge color as it is not fully reliable. Phase-resolved spectroscopy of \emph{NICER} data confirms this trend, albeit with fewer phase bins and more uncertain constraints (see fitted temperatures in Table~\ref{tab:spec}). The trends shown in Fig.~\ref{fig:energy_evolution} are in agreement with other known QPE sources. This makes the QPE classification of J0325 robust.

\begin{figure}[t]
		\centering
		\includegraphics[width=0.95\columnwidth]{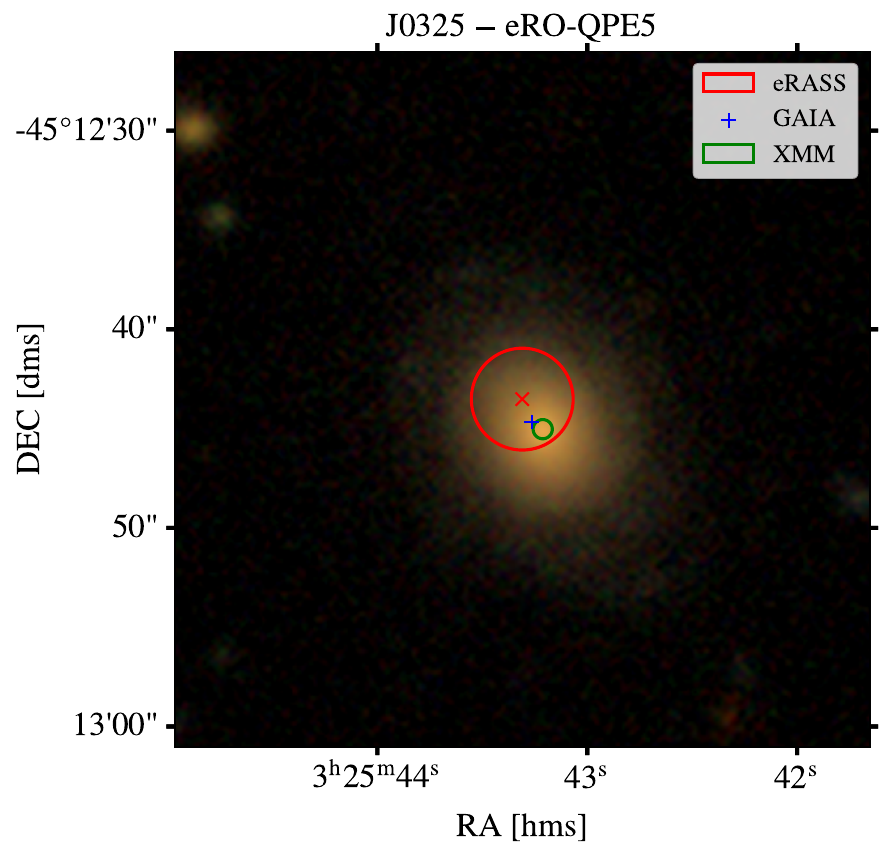}
    	\caption{$35$"$\times\,35$" cutout of the DESI Legacy Imaging Surveys Data Release 10 [Legacy Surveys / D. Lang (Perimeter Institute)] with the X-ray 1$\sigma$ position circles in red (eROSITA) and green (\emph{XMM-Newton}).}
		\label{fig:opt_image}
\end{figure}

\begin{figure*}[t]
		\centering
        \includegraphics[width=0.9\textwidth]{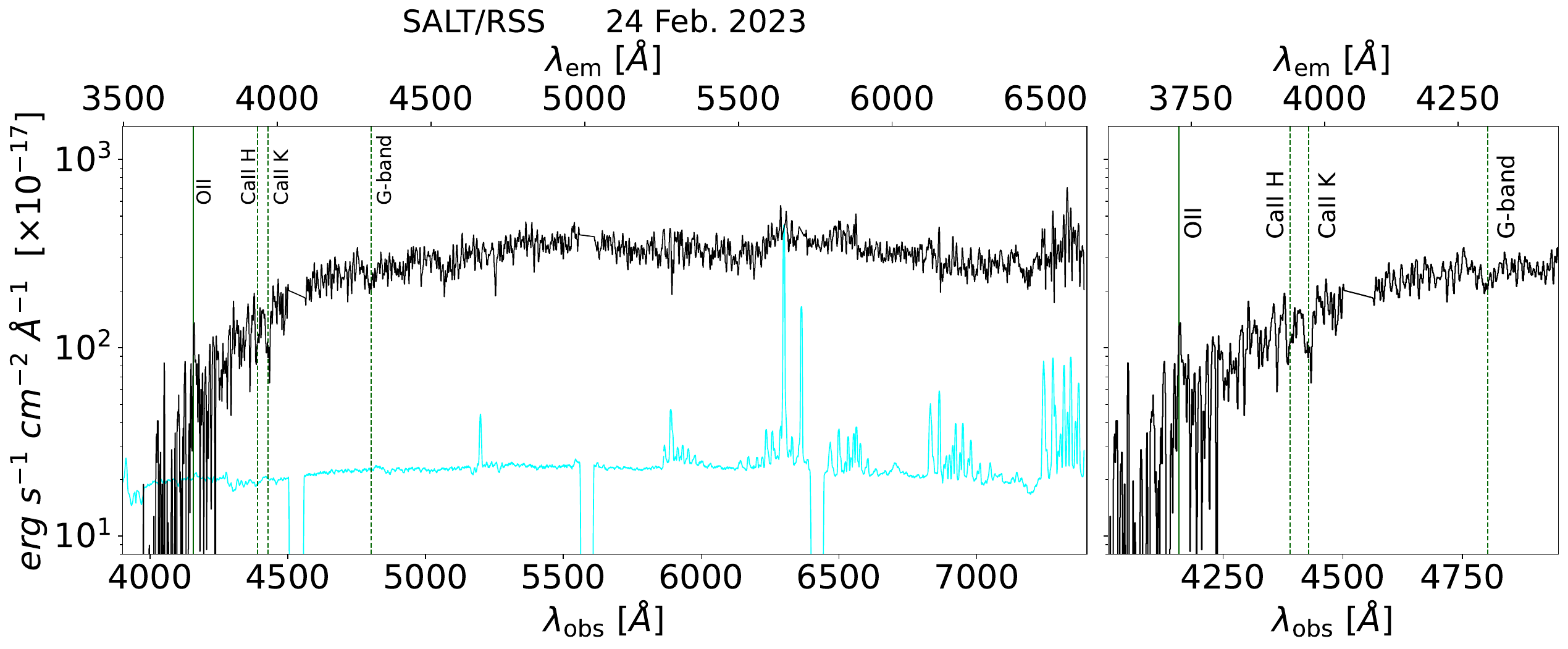}
    	\caption{SALT/RSS optical spectrum of eRO-QPE5 with a zoom-in around the Calcium absorption doublet that indicates a spec-z of $0.1155$. The sky spectrum is shown in cyan in arbitrary units.}
		\label{fig:opt_spec}
\end{figure*}

\section{Multiwavelength emission}
\label{sec:multiwav}

\subsection{Optical-UV-IR}
\label{sec:opt}

eRO-QPE5 is associated with a galaxy at (RA, Dec) = (03:25:43.21250, -45:12:45.1286) detected in the Dark Energy Survey Data Release 2 \citep{Abbott:2021:DES}. We show the Legacy Survey DR10 optical image in Fig.~\ref{fig:opt_image} with the eROSITA X-ray position (and 1$\sigma$ accuracy) shown with a red cross and circle (from the cumulative eRASS:4 image). The more accurate \emph{XMM-Newton} localization is shown in green, indicating that eRO-QPE5 is consistent with the nucleus given the current positional accuracy. Infrared photometry from NEOWISE \citep{Mainzer+2014:neowise} shows galaxy-like colors ($\rm W1- \rm W2\sim0$, i.e., compatible within uncertainties), as opposed to AGN-like colors, for the last twelve years, with no significant variability with the probed six-month cadence. No significant variability nor obvious flares are present in any of the archival ASASSN \citep{Shappee+2014:ASASSN,Kochanek+2017:asassn} and ATLAS \citep{Tonry+2018:atlas} optical light curve databases, going back about 12 and 8 years, respectively. Its optical spectrum appears rather featureless and noisy, as observed by both SALT/RSS and Magellan/MagE, although Calcium absorption lines and weak [O$_{\rm II}$] in emission allowed us to estimate $z=0.1155$. We show the SALT spectrum in Fig.~\ref{fig:opt_spec}, while the more uncertain MagE spectrum in Sect.~\ref{sec:app_opt} (together with details on data processing). This inferred spec-z would make eRO-QPE5 the most distant QPE source to date. 

\begin{figure}[tb]
		\centering
		\includegraphics[width=0.9\columnwidth]{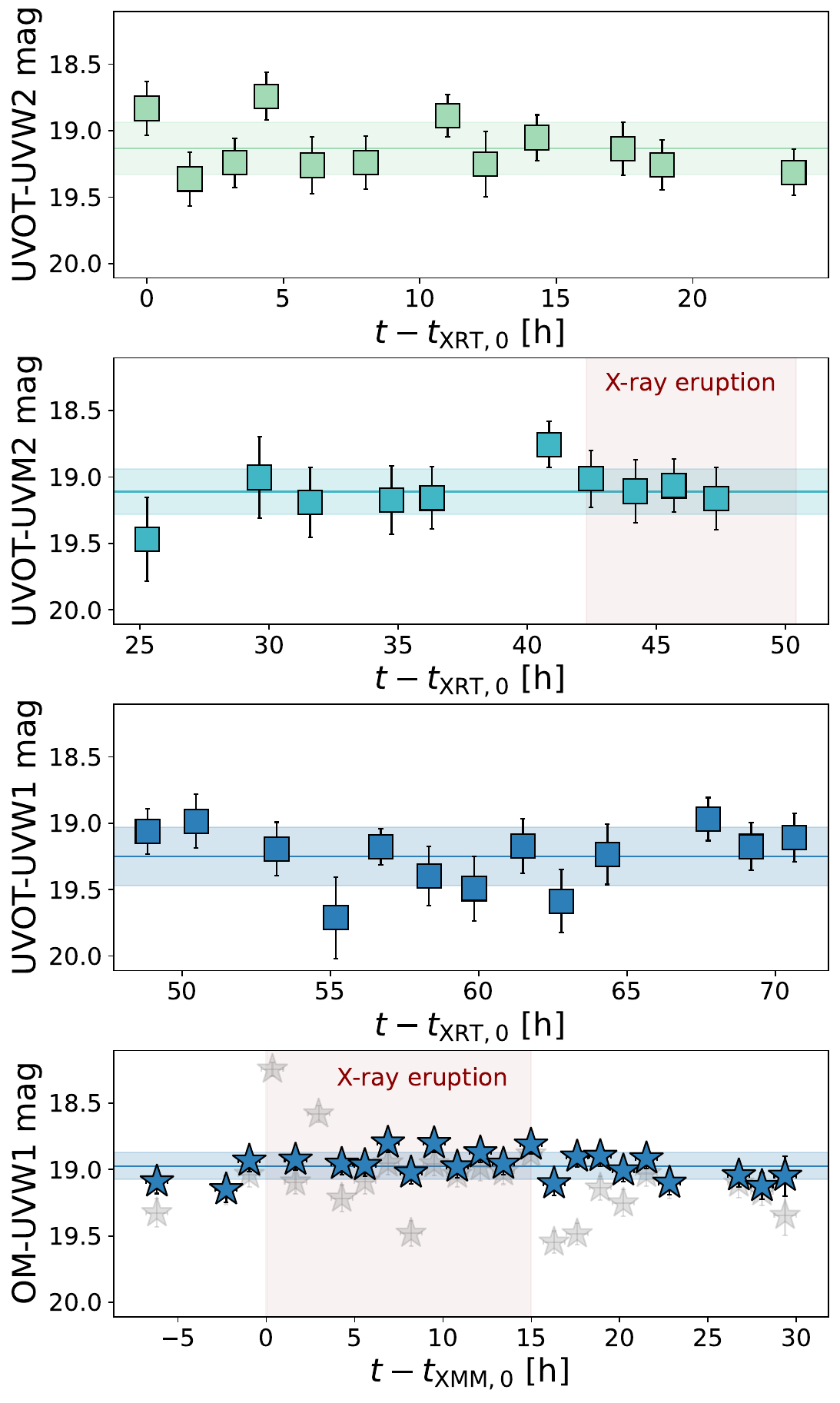}
		\caption{UV light curves of eRO-QPE5 taken with, from top to bottom, the \emph{Swift}/UVOT UVW2 (208.4\,nm), UVM2 (224.5\,nm), and UVW1 (268.2\,nm) filters, and the UVW1 (291\,nm) of the OM aboard \emph{XMM-Newton}. UVOT data are scaled at the start of the UVW2 exposure, while OM data are scaled at the start of the X-ray exposure. The OM panel shows also the uncorrected noisier light curve from the standard \texttt{omichain} task (gray, see Sect.~\ref{sec:sed} for more details). The times of the X-ray eruptions observed by XRT and \emph{XMM-Newton} are shown with the red shaded region. In all panels, the mean magnitude and its standard deviation are highlighted with the same color coding.}
		\label{fig:om}
\end{figure}

\emph{Swift}/UVOT and \emph{XMM-Newton} Optical Monitor (OM) photometry was taken in simultaneity with X-rays. The top three panels of Fig.~\ref{fig:om} show \emph{Swift}/UVOT light curves with the UVW2 (208.4\,nm), UVM2 (224.5\,nm), and UVW1 (268.2\,nm) filters, from top to bottom, respectively, scaled at the start of the UVW2 exposure and showing only detections above $3\sigma$. The related mean standard deviation values are $19.25 \pm 0.22$, $19.11 \pm 0.17$, and $19.13 \pm 0.20$, for the UVOT UVW1, UVM2, UVW2 filter, respectively. The bottom panel shows OM UVW1 data taken at 291\,nm, which are also constant with mean AB magnitude of $18.97$ with a standard deviation of 0.10, compatible with the average uncertainties in the individual measurements of $\sim0.08$. The times of the X-ray eruptions observed by XRT and \emph{XMM-Newton} are shown with red shaded regions, confirming the absence of significant simultaneous flaring variability in the UV, as reported before from OM and UVOT data \citep[e.g.,][]{Arcodia+2021:eroqpes,Arcodia+2024:eroqpes}. This finding was recently confirmed at higher confidence with Hubble Space Telescope (HST) data of eRO-QPE2 \citep{Wevers+2025:sed}, which ruled out UV variability at a flux level that is $\sim100$ times fainter than what it typically probed by OM or UVOT data. The detected constant point source at a luminosity level of $\lambda L_{\rm UV}\sim$\,few$\times10^{41}\,$erg\,s$^{-1}$ is dominated by the quiescent accretion disk \citep{Wevers+2025:sed}. In comparison, if the accretion disk in eRO-QPE5 were as UV luminous as that of eRO-QPE2 (which is a fair assumption given the comparable soft X-ray luminosity of the two disks), it would correspond to a $159.6\,$nm flux of $\approx0.75\,\mu$Jy, and to even lower fluxes at longer wavelengths given the steep accretion disk SED. Thus, it would be $>25$ fainter than the observed OM and UVOT fluxes of $\approx0.02\,$mJy in the $208-268$\,nm range, suggesting that the latter flux estimates are dominated by the galaxy. This lack of constraining power for the nuclear emission is perhaps unsurprising given that HST photometry was obtained with a 0.122\,arcsec aperture, compared to the 5.7 and 5\,arcsec used for the OM and UVOT, respectively.

\begin{figure}[tb]
		\centering
		\includegraphics[trim={0 0 1.5cm 1.4cm},clip,width=0.99\columnwidth]{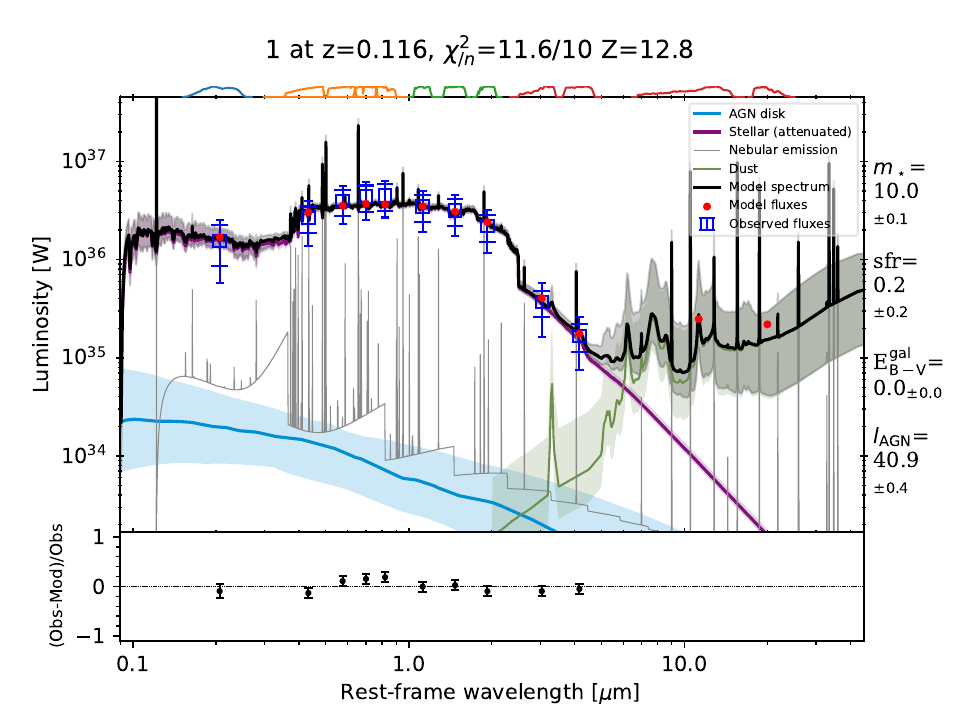}
    	\caption{Spectral energy distribution of eRO-QPE5, with the GRAHSP UV-optical-IR SED fitting with galaxy and AGN templates, with the former (purple and green) dominating on the latter (light blue). The fitted $M_*$ is $8.5^{+3.3}_{-1.7}\times10^9\,M_{\astrosun}$. 
        }
		\label{fig:sed}
\end{figure}

We performed SED photometry fitting to obtain an estimate of $M_*$. We collated Legacy Survey DR10 photometry with \texttt{RainbowLasso}\footnote{\href{https://github.com/JohannesBuchner/RainbowLasso}{https://github.com/JohannesBuchner/RainbowLasso}} \citep{Buchner+2024:GRAHSP} and fit the SED with the Genuine Retrieval of AGN Host Stellar Population (GRAHSP; \citealp{Buchner+2024:GRAHSP}) code. The models used in GRAHSP include AGN components (continuum, emission lines, torus) and galaxy components, both attenuated by dust and redshifted. For the stellar component, we adopted the stellar population synthesis \texttt{m2005} \citep{Maraston+2005:sps} combined with the parametric tau-delayed star formation history \texttt{sfhdelayed} \citep[see][]{Buchner+2024:GRAHSP}. Given the known unusual shape of the accretion disk SED in QPE sources compared to canonical AGN (i.e. no broad lines, no torus, X-ray corona, radio emission), we only include the continuum for the AGN component, and given the low column density observed in the X-rays we set the nuclear obscuration to zero to mitigate possible degeneracies with the continuum. These differences have to be kept in mind for the rest of the manuscript. While the nuclear component is referred to as AGN to follow the language of the code and the SED fitting literature, the ``AGN component" is here only a disk, and it is very different from extended AGN disks \citep[e.g.,][]{Nicholl+2024:qiz,Wevers+2025:sed,Guolo+2025:sed}. GRAHSP parametrizes the accretion disk as a smooth bending power-law, and we leave the break wavelength free to vary between $\sim10-30$\,nm to roughly match the peak of the SED to the soft X-rays. The normalization of the AGN disk continuum is the $5100\,\AA$ monochromatic luminosity and we tested different priors, including a uniform one and a broad Gaussian prior centered at $L_{\rm UV}\sim 10^{41}$erg\,s$^{-1}$. We adopt the latter for the final inferred parameters, and show it in Fig.~\ref{fig:sed} (and Fig.~\ref{fig:sedmJy} in flux units). In either case, we note that the fitted AGN disk continuum is subdominant compared to the galaxy (in agreement with the reasoning above of OM/UVOT data being dominated by the galaxy), thus the outer radius truncation or lack thereof does not impact any of the results, and the AGN disk luminosity prior only affects the posterior edges (see Sect.~\ref{sec:sed}). This estimated and fitted \text{``AGN component''} level also roughly matches that of the observed UV disk emission in eRO-QPE2 \citep[e.g.,][]{Wevers+2025:sed} and it is also within the typical range of $L_{\rm UV}$ in TDE accretion disks (A. Mummery, priv. comm.; \citealp{Mummery+2024:scaling}). With this model setup, we obtained $M_* = 8.5^{+3.3}_{-1.7}\times10^9\,M_{\astrosun}$. Adopting widely-used scaling relations between black hole and total stellar mass of the galaxy \citep{Reines+2015:mstar}, including uncertainties and intrinsic scatter of the relation, we obtain $M_{\rm BH}=2.9^{+5.4}_{-2.2}\times10^7\,M_{\astrosun}$. Thus, eRO-QPE5 sits at the upper bound of what has been discovered so far in QPEs with eROSITA, together with eRO-QPE4 \citepalias{Arcodia+2021:eroqpes,Arcodia+2024:eroqpes}.

\subsection{Radio non-detections}
In order to search for transient radio emission associated with ejections from the QPEs or non-transient host galaxy radio emission, we observed the coordinates of eRO-QPE5 with the Australia Telescope Compact Array (ATCA) on 
2024 April 26 (proposal ID C3586), which is not in simultaneity with any X-ray observation. We observed using the dual 4\,cm receiver with the 2x2\,GHz bandwidth centered on 5.5 and 9\,GHz respectively and 2048 spectral channels in each band. The observation was 4\,hr long allowing for approximately 200\,min of integration time on the target field. The array was in the extended 6A configuration allowing an image resolution of approximately 3" to be achieved. All data were reduced in the Common Astronomy Software Package \citep[CASA;][]{CASATeam2022} following standard procedures. PKS 1934--638 was used for flux and bandpass calibration and PKS 0244--470 was used for phase calibration. Images of the target field were created using the CASA task \texttt{tclean}. No radio source was detected at the coordinates of eRO-QPE5 with a 3$\sigma$ upper limit of $<$36$\mu$Jy at 5.5\,GHz and $<$33$\mu$Jy at 9\,GHz. 


\section{Correlation Analysis and comparisons with collision models}

Given the growing population of confirmed QPE sources, we now attempt to find some possible correlations to test current and future models. Naturally, some quantities are more easily measurable than others. For instance it is relatively easy to infer QPE duration, recurrence and peak temperatures for nearly all sources with little model dependence, while it is more complicated to infer the properties of the much fainter quiescence, and source properties such as black hole mass and accretion rate in a systematic way in lack of homogeneous multi-band high-resolution and high signal-to-noise data. In this work, we collated QPE durations and recurrences from the literature adopting, or converting to, a rise-to-decay (or $3\sigma$ interval) for the QPE duration. In addition, to overcome biases in using different $M_{\rm BH}$ estimates, we collected $M_{\rm BH}$ from the literature \citep{Sun+2013:rxj,Shu+2017:rxj,Arcodia+2021:eroqpes,Arcodia+2024:eroqpes,Wevers+2022:host,Wevers+2024:host,Wevers+2025:sed,Nicholl+2024:qiz,Guolo+2025:sed,Newsome+2024:upj,SanchezSaez+2024:ansky,Chakraborty+2025:upj} and averaged over possible different methods (e.g., velocity dispersion, stellar mass, disk fitting) when available, adding all uncertainties in quadrature together with a $0.5\,$dex systematic. This results in a mean $1\sigma$ uncertainty of $\sim0.72$\,dex, which, for reference, is only $\sim2-3$ times smaller than the dynamic range probed by QPE $M_{\rm BH}$ values. Intuitively, this hampers any attempt to find significant correlations with $M_{\rm BH}$ values (see the Sections below). The values adopted in this work are reported in Table~\ref{tab:mbh}. Regarding quiescence values, we averaged over multiple epochs, if available, to account for possible variability and scatter. We collected from the literature quiescence values with the simplistic redshifted \texttt{diskbb} in most cases, except AT2019qiz, eRO-QPE2, and GSN\,069 for which a broadband disk fit was performed \citep{Nicholl+2024:qiz,Wevers+2025:sed,Guolo+2025:sed}. We acknowledge that this may be source of biases within a factor $\sim2$, and that future work may improve on this approach. For eRO-QPE3, we use the eRASS1 and eRASS2 quiescence estimates, and \emph{XMM-Newton} fit values for all others. Since in some sources $kT_{\rm QPE}$ may vary significantly \citep[e.g.,][]{Giustini+2024:rxj}, and eruptions may appear colder when fainter \citep[e.g.,][]{Miniutti+2023:alive}, for sources with a single characteristic peak temperature reported and obvious differences in eruption amplitudes we also added a $5\,$eV systematic in quadrature to tentatively account for epoch-to-epoch dispersion. 

Throughout this section, we report fits done by drawing samples of the various quantities (in log-log space) from a normal distribution centered at the mean value (reported in this paper or the literature) and using their $1\sigma$ uncertainties as standard deviation. We then fit a line with Gaussian scatter while marginalizing over the samples for each data point. 
This accounts for the entire distribution of data and not just fitting $1\sigma$ uncertainties. In case of asymmetric uncertainties, we average over the positive and negative for the purpose of fitting correlations. All quantities are scaled to the rest frame. The fit is performed with the \texttt{Ultranest} package \citep{Buchner2021:ultranest} which implements  MLFriends \citep{Buchner2016,Buchner2019}, a nested sampling algorithm \citep{Skilling2004,Buchner2023}.

\subsection{Correlations between QPE timescales, and with $M_{\rm BH}$}
\label{sec:correlations}

\begin{figure}[tb]
		\centering
		\includegraphics[width=0.99\columnwidth]{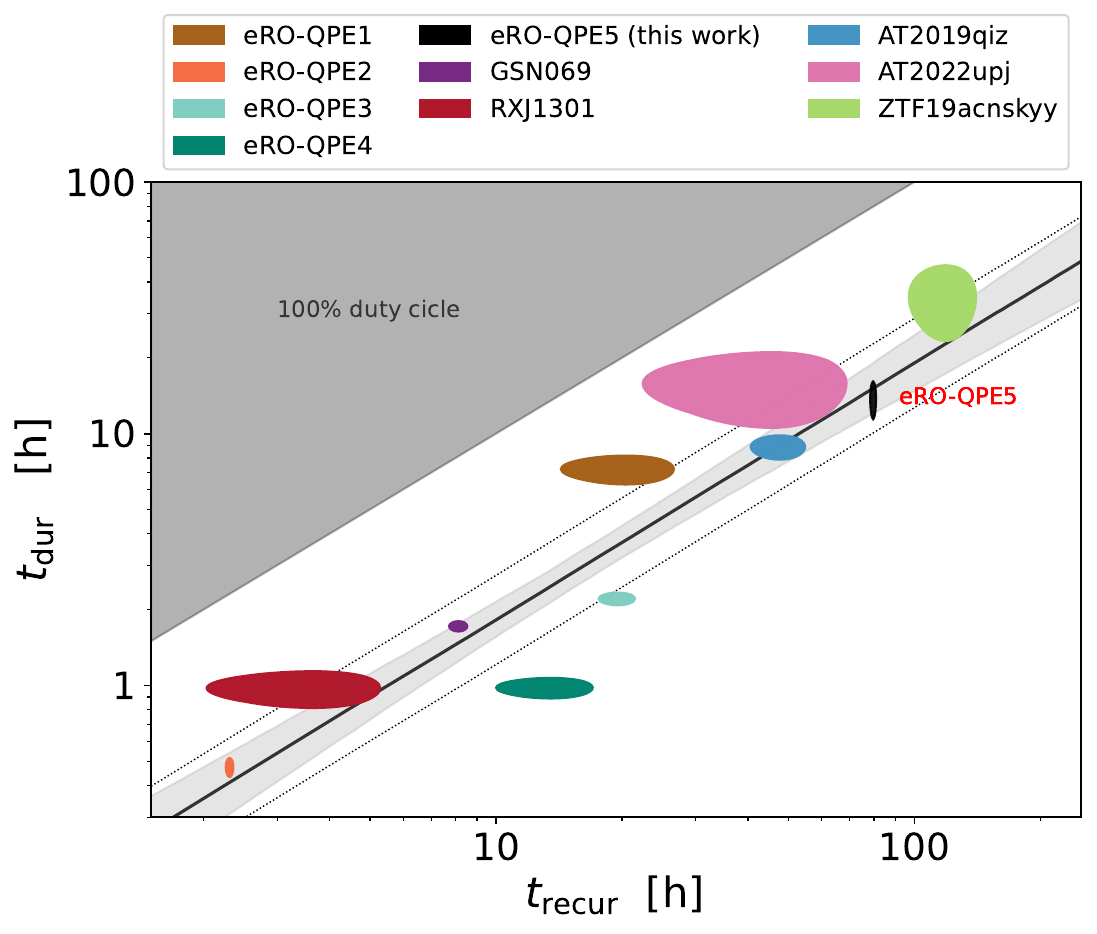}
		\caption{Correlation between QPE rest-frame duration ($t_{\rm dur}$) and recurrence time ($t_{\rm recur}$). We show the fitted median model (solid black line) and related $1\sigma$ contours (shaded black area), with a slope of $1.01\pm0.12$ and an intrinsic scatter of $0.18 \pm 0.08$ (dotted black line). As pointed out in \citet{Chakraborty+2025:upj}, data are forced to be below the one-to-one line representing duty cycles of $100\%$ (grey wedge), and current instruments are biased against finding sources with short $t_{\rm dur}$ and long $t_{\rm recur}$ (i.e., the bottom right corner). }
		\label{fig:tdurtrecur}
\end{figure}

An apparent correlation between QPE duration and recurrence was pointed out in earlier works with the first handful of QPE sources or candidates \citep{Chakraborty+2021:qpe5cand,Guolo+2023:swift,Arcodia+2024:eroqpes}, together with other possible correlations (e.g., with black hole mass, luminosity) which seemed however either ruled out or remained ambiguous. The latest QPE discoveries \citep{Nicholl+2024:qiz,Chakraborty+2025:upj,Hernandez-Garcia+2025:ansky} happened to confirm this $t_{\rm dur} - t_{\rm recur}$ correlation by extending it to longer timescales, including eRO-QPE5 from this work. We show this in Fig.~\ref{fig:tdurtrecur} for the sources with at least two full consecutive eruptions and which show the characteristic energy evolution (e.g., Fig.~\ref{fig:energy_evolution}). Apart from possible differences mainly due to the definition of QPE duration, the observed slope appears to be $\approx 1$ \citep{Chakraborty+2025:upj,Hernandez-Garcia+2025:ansky}. We fitted the data in Fig.~\ref{fig:tdurtrecur} using a linear model and obtained a slope of $1.01\pm0.12$, shown as a median (solid black line) and related $1\sigma$ contours (shaded black area), with an intrinsic scatter or $0.18 \pm 0.08$\,dex (dotted black line). In other words, QPEs seem to be observed at a constant duty cycle \citep[e.g.,][]{Guolo+2023:swift,Nicholl+2024:qiz}, which from Fig.~\ref{fig:tdurtrecur} we constrain at $\sim18\%$ (using rise-to-decay durations as in the plot, the duty cycle would naturally decrease taking near-peak durations). As pointed out by \citet{Chakraborty+2025:upj}, data are forced to be below the one-to-one line representing duty cycles of $100\%$ (shown as a wedge in Fig.~\ref{fig:tdurtrecur}), thus a slope much larger than one is unlikely to ever be found, and there are poorly-investigated limitations in discovering QPEs with short durations and long recurrence times (the lower right corner), which may artificially steepen the correlation.

\begin{figure}[tb]
		\centering
		\includegraphics[width=0.99\columnwidth]{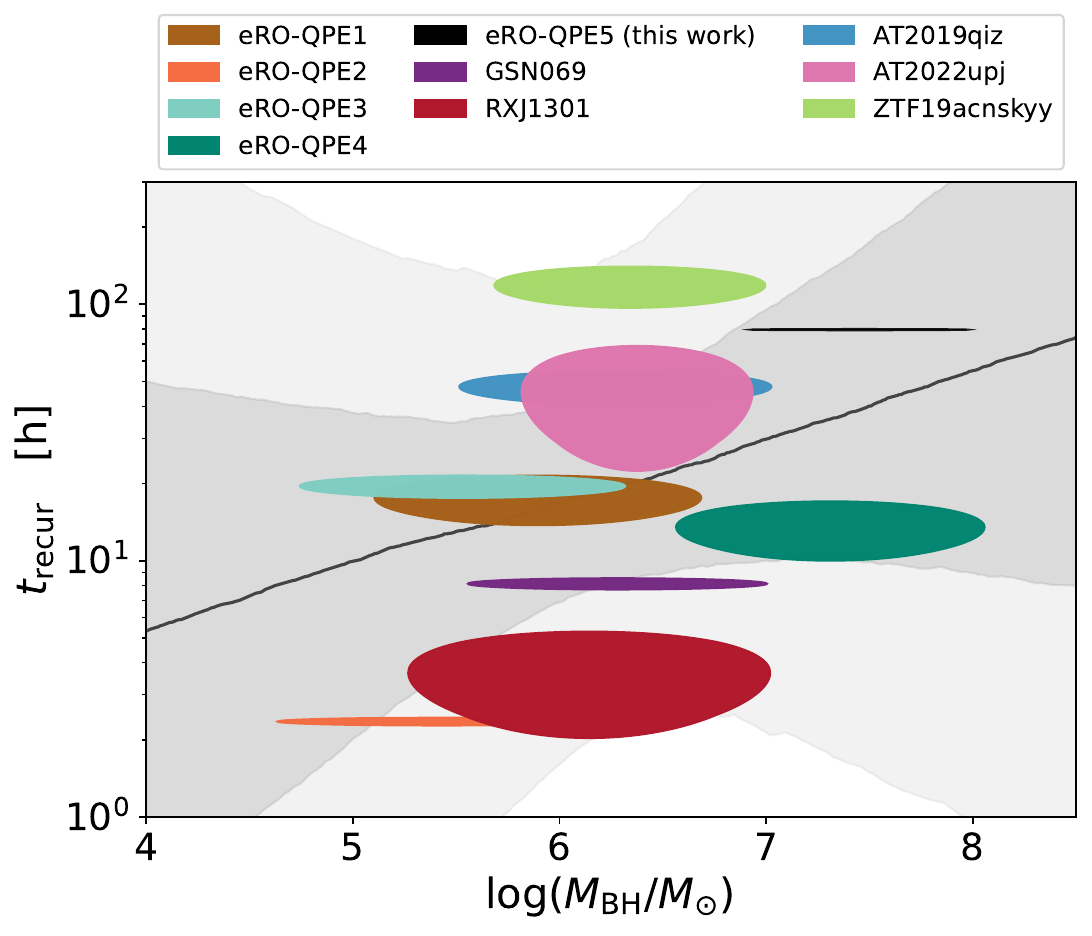}
		\includegraphics[width=0.99\columnwidth]{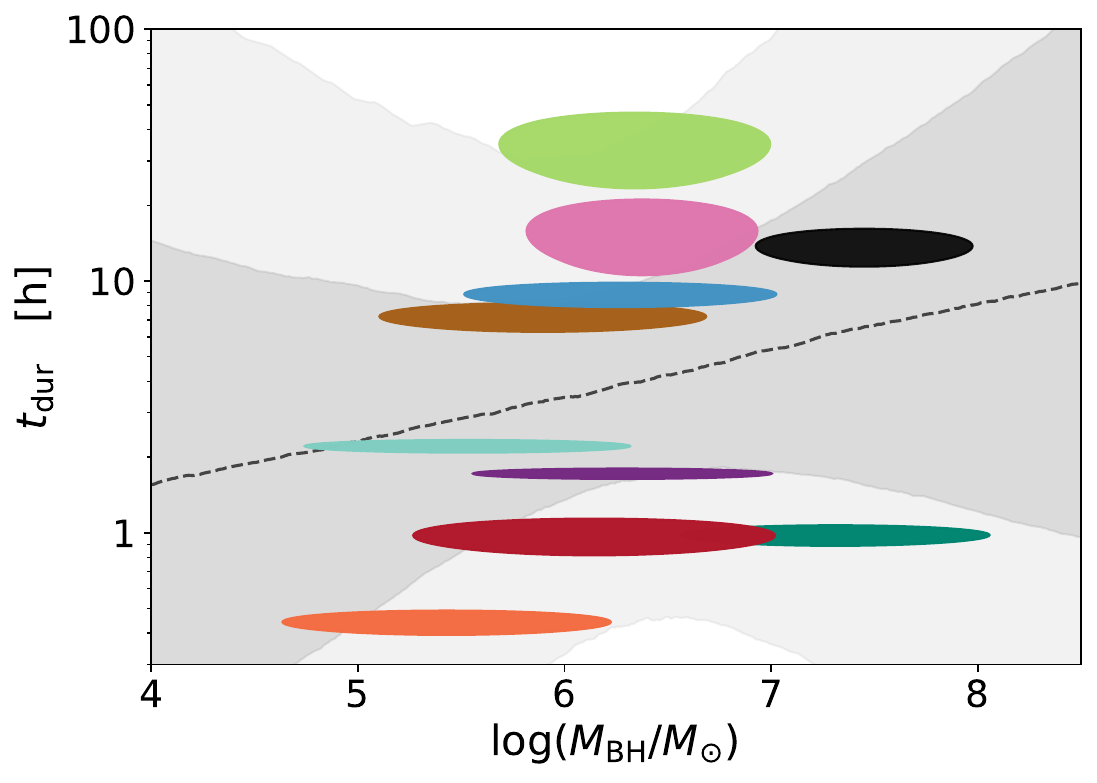}
    	\caption{Lack of correlation between $t_{\rm recur}$ and $M_{\rm BH}$ in the top panel, and between $t_{\rm dur}$ and $M_{\rm BH}$ in the bottom panel. In both cases, the fitted slope is consistent with zero within $1\sigma$ (we show $1\sigma$ and $3\sigma$) and the linear term is not statistically required.}
		\label{fig:tdurmass}
\end{figure}

Previous works have reported a correlation between $t_{\rm recur}$ and $M_{\rm BH}$ \citep{Zhou+2024:longterm}, while others have not found any \citep{Nicholl+2024:qiz}. This may be due to the small number statistics, and most likely to the highly uncertain black hole masses and the various methods used to infer them. In this work, we overcome possible biases by averaging over different methods. We did not find any significant correlation between neither $t_{\rm recur}$ and $M_{\rm BH}$, nor $t_{\rm dur}$ and $M_{\rm BH}$ (Fig.~\ref{fig:tdurmass}). In both cases, the linear fit yields a slope consistent with zero at $1\sigma$ level, and the linear term is not statistically required with respect to a constant. This has more to do with the statistical and systematic uncertainties in inferring $M_{\rm BH}$ in low-mass galaxies, rather than any intrinsic correlation, or lack thereof. We consider our conservative approach of averaging across methods and adding $\sim0.5\,$dex of systematics crucial to avoid over-interpreting possible correlations with $M_{\rm BH}$. However, we also tested that our conservative approach does not drive the absence of a significant correlation simply by increasing uncertainties. We tested the same correlations using only $M_{BH}$ values from either TDE disk fitting (for eRO-QPE2, GSN\,069, AT2019qiz and AT2022upj; \citealp{Wevers+2025:sed,Guolo+2025:sed,Nicholl+2024:qiz,Chakraborty+2025:upj}) or from the $M_{BH}-\sigma$ relation inferred with high-resolution MagE or MUSE spectra \citep[for RXJ1301, eRO-QPE1 and eRO-QPE3;][]{Wevers+2022:host,Wevers+2024:host}. The average uncertainty of these $M_{\rm BH}$ estimates is $\sim0.34\,$dex, less than half the one reported in Table~\ref{tab:mbh}. No significant correlations is found between neither $t_{\rm recur}$ and $M_{\rm BH}$, nor $t_{\rm dur}$ and $M_{\rm BH}$, confirming the results obtained previously.

\begin{table}[tb]
	\small
    \centering
	\setlength{\tabcolsep}{10pt}
	\renewcommand{\arraystretch}{1.1}
	\caption{Black hole masses used in this work.}
	\label{tab:mbh}
	\begin{threeparttable}
		\begin{tabular}{ccc}
			\toprule
            \multicolumn{1}{c}{Source} &
			\multicolumn{1}{c}{$\log M_{\rm BH}$} &
			\multicolumn{1}{c}{Ref.}
            \\
            \midrule
             eRO-QPE1 &  $5.90 \pm 0.79$  &   [1-2]    \\
             eRO-QPE2 &  $5.43 \pm 0.79$   &  [1-3]  \\
             eRO-QPE3 &  $5.53 \pm 0.79$   &    [4-5]   \\
             eRO-QPE4 &  $7.31 \pm 0.75$   &    [4]   \\
             eRO-QPE5 &  $7.45 \pm 0.52$   &    this work   \\
             GSN 069 &   $6.28 \pm 0.72$  &    [2, 6]   \\
             RX J1301 &   $6.14 \pm 0.88$  &    [5, 7, 8]   \\
             AT2019qiz &   $6.27 \pm 0.76$  &   [9]    \\
             AT2022upj &   $6.38 \pm 0.56$  &    [10,11]   \\
             ZTF19acnskyy &   $6.34 \pm 0.66$  &   [12]    \\
             \bottomrule
		\end{tabular}
        \begin{tablenotes}
        \small
        \item Values are averaged from the listed references using various methods (from scaling with stellar mass, velocity dispersion, SED fitting), and their uncertainties are added in quadrature with a $0.5\,$dex systematic. References correspond to \citet{Arcodia+2021:eroqpes,Wevers+2022:host}; \citet{Wevers+2025:sed,Arcodia+2024:eroqpes,Wevers+2024:host,Guolo+2025:sed,Sun+2013:rxj,Shu+2017:rxj,Nicholl+2024:qiz,Newsome+2024:upj}; \citet{Chakraborty+2025:upj}; \citet{SanchezSaez+2024:ansky}; ordered from [1] to [12], respectively.
        \end{tablenotes}
   \end{threeparttable}
\end{table}

\subsubsection{Comparisons with theoretical predictions from collision models}

We compare these observed correlations (or lack thereof) with predictions from the disk collision theory \citep[e.g.,][]{Linial+2023:qpemodel,Franchini+2023:qpemodel}. If $t_{\rm recur}$ is a good proxy for the orbital period, it should scale as $t_{\rm recur} \propto M_{\rm BH}$, while we do not find any correlation here (Fig.~\ref{fig:tdurmass}). We note that the original prescriptions are not able to reproduce many observables, as also pointed out in very recent work emerged during the final stages of our manuscript \citep[e.g.,][]{Guolo+2025:gsnlongterm,Guo+2025:coll,Mummery+2025:collisions,Linial+2025:streams}. In the original prescriptions, if $t_{\rm dur}$ is a proxy for the diffusion time of the expanding gas bubble \citep[e.g., eq.\ 16 in][]{Linial+2023:qpemodel}, the theory predicts a dependence $t_{\rm dur} \propto t_{\rm recur}^{2/3}$. This prediction can be qualitatively interpreted as follows: everything else being the same, collisions at larger mass-normalized distances from the black hole (i.e., higher $t_{\rm recur}$) would occur at lower relative velocities between the orbiter and the disk (thus lower shock velocities and higher diffusion times $\sim t_{\rm dur}$ in the expanding gas bubble, e.g., eq.\ 16 in \citealp{Linial+2023:qpemodel}). While the predicted slope is lower than the observed slope $\sim1$ (Fig.~\ref{fig:tdurtrecur}), a few theoretical uncertainties hamper a direct comparison between these two dependencies, in addition to the observational biases discussed above. First, the theoretical prediction assumes that all disks are close to standard $\alpha$-disks \citep[e.g.,]{Shakura+1973:alphadisk} with a similar $\alpha$ value constant with radius, and are all accreting at a similar fraction $\dot{m}$ of the Eddington limit. 
Similarly, the orbiter's cross section is assumed close to the radius of the Sun for all sources. The assumption is thus that none of these  are a function of radius, and only impact the scatter. Second, $t_{\rm dur} \propto \kappa^{1/2}$ where the ejecta opacity $\kappa$ corresponds to that of the disk at the collision radius, for high enough inclinations \citep{Linial+2023:qpemodel}. While $\kappa$ is typically assumed constant at the Thomson value, in fact it may be a strong function of radius depending on the dominating process and including the iron bump \citep[e.g.,][]{Jiang+2016:iron,Jiang+2020:iron}, which becomes a relevant contributor at the disk temperatures and densities expected for the black hole masses of QPE sources. To test the impact of the $\kappa (r)$ dependency, we ran a few realizations of the accretion disk model presented in \citet{Arcodia+2019:diskcorona}, which is based on $\alpha$ disks with the self-consistent addition of energy lost from the disk to the corona and of alternative viscosity prescriptions, and which includes accurate opacity tables from \citet{Seaton+1994:OP} instead of just the electron-scattering opacity\footnote{\href{https://github.com/rarcodia/DiskCoronasim}{https://github.com/rarcodia/DiskCoronasim}}. 
Adopting $\log (M_{\rm BH}/M_{\odot}) = (5, 6, 7)$, relevant for QPE sources, we find that at all relevant radial distances (between $\approx50-1000$ gravitational radii) $\kappa$ is always two to several times larger than Thomson and depends strongly on radius (e.g., see Fig. A1 in \citealp{Arcodia+2019:diskcorona}). Thus, the radial dependency from opacity needs to be included to isolate the dependency from the QPE period alone. Furthermore, the diffusion time also depends on black hole mass as $t_{\rm dur} \propto M_{\rm BH}^{-2/3}$, although this is however not found with current data (bottom panel of Fig.~\ref{fig:tdurmass}). We also tested whether the combined dependency $t_{\rm dur} \propto t_{\rm recur}^{2/3} \, M_{\rm BH}^{-2/3}$ may be recovered in a 3D linear fit. We obtained a slope of $1.16 \pm 0.18$ and $-0.20 \pm 0.13$ for $t_{\rm recur}$ and $M_{\rm BH}$, respectively. Thus, the 3D fit is inconsistent with the diffusion time dependency, and the coefficients inferred from the separate 2D relations ($\sim1$ for $t_{\rm recur}$ and $\sim0$ for $M_{\rm BH}$) are recovered within $1\sigma$ uncertainties. 
Furthermore, recently \citet{Mummery+2025:collisions} pointed out that in case of a TDE origin for all the QPE sources' accretion disks, the steady-state assumption should not be used.

Finally, we note that, in the case of stellar orbiters, hydrodynamic simulations have shown that QPEs may actually be powered by collisions between stellar debris liberated in previous collisions and the accretion disk \citep{Yao+2025:simul}. Thus, the QPE duration would not be set by the diffusion time in this case, but rather by the dispersion in arrival time of the stellar
debris in front of and behind the star. Assuming for simplicity that the collisions are dominated by debris inside the Hills sphere \citep[][their Eq.~14]{Yao+2025:simul}, the QPE duration would scale as $t_{\rm recur} \, M_{\rm BH}^{-1/3}$, which is tantalizingly close to the observed dependencies of the 3D fit. This interpretation is also in agreement with recent work finding that the flares' integrated energy cannot be reproduced by the disk mass if swept by the physical size of the star \citep{Guo+2025:coll}. Indeed, \citet{Mummery+2025:collisions} pointed out that the only way to reconcile the high integrated energy found in some QPE sources with TDE disks \citep{Nicholl+2024:qiz,Chakraborty+2025:upj,Guolo+2025:gsnlongterm} is to have the collisions be powered by stellar debris, as suggested by the simulations in \citet{Yao+2025:simul} and by \citet{Linial+2025:streams}. We note that also eRO-QPE5 shows large integrated energies ($\sim3.4 \times 10^{47}$\,erg) and would thus require the orbiter to be a puffed up and elongated star to reproduce the observed eruptions, if collisions on a similar TDE-like disk are in place.

\subsection{Correlations with disk and QPE temperatures}
\label{sec:correlations2}

\begin{figure}[tb]
		\centering
		\includegraphics[width=0.99\columnwidth]{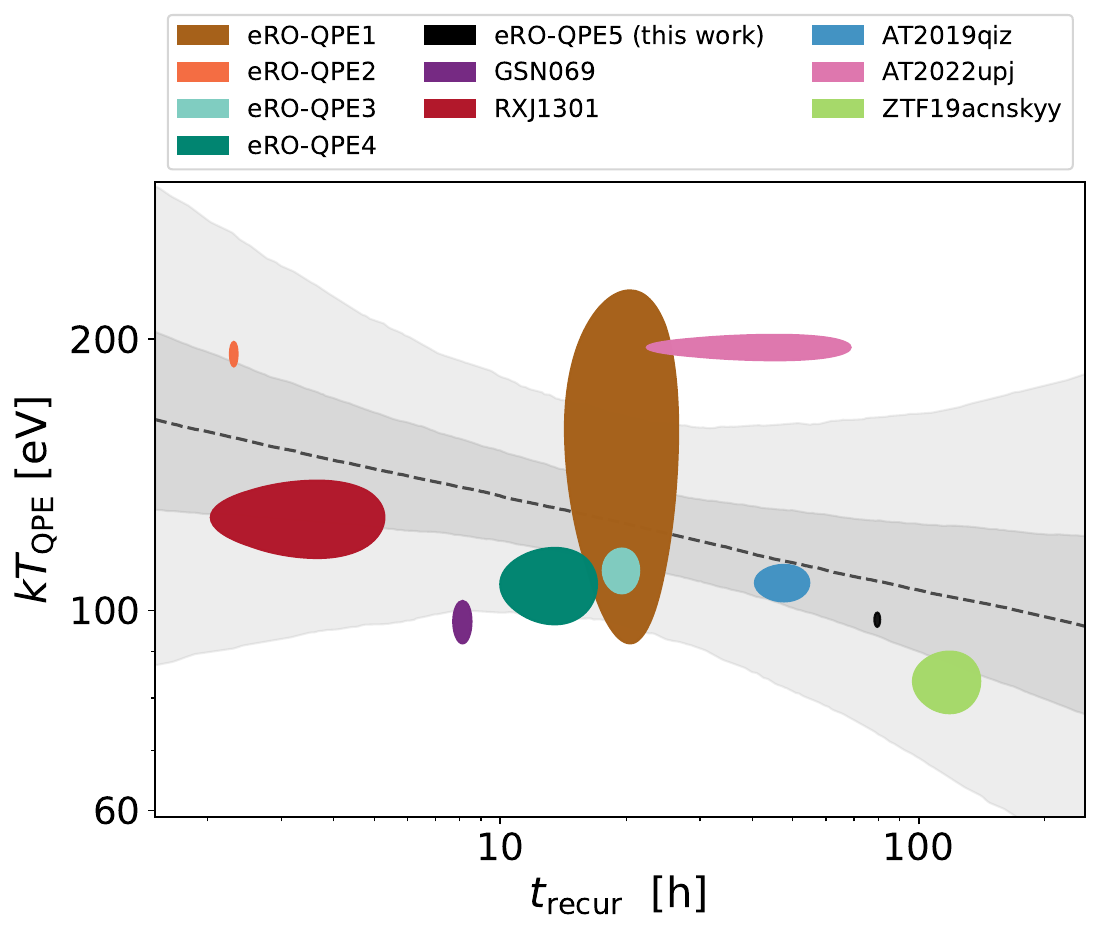}
    	\caption{Lack of correlation between $t_{\rm recur}$ and $kT_{\rm QPE}$. The slope is consistent with zero within $2\sigma$ (we show $1\sigma$ and $3\sigma$) and the linear term is not statistically required.}
		\label{fig:kt_trec}
\end{figure}

Here we investigate possible correlations between timing properties ($t_{\rm dur}$ or $t_{\rm recur}$) and spectral quantities, such as the quiescent disk temperature $kT_{\rm disk}$, the peak QPE temperature $kT_{\rm QPE}$, and the bolometric disk luminosity. We do not find any significant correlation between $t_{\rm dur}-kT_{\rm QPE}$, $t_{\rm recur}-kT_{\rm QPE}$ (e.g., Fig.~\ref{fig:kt_trec}), nor between $t_{\rm dur}-kT_{\rm disk}$ and $t_{\rm recur}-kT_{\rm disk}$. In the $t_{\rm recur} - kT_{\rm QPE}$ case, assuming a disk-collision scenario, a lack of correlation would suggest that $kT_{\rm QPE}$ does not strongly depend on the underlying disk at the collision point, since $t_{\rm recur}$ is a function of the collision radius for a given black hole mass. In addition, neither $kT_{\rm disk}$ nor $kT_{\rm QPE}$ appear to correlate with $M_{\rm BH}$. In the former case, this is at first glance puzzling given the accretion disk origin for the quiescent X-ray emission. We speculate the cause being the $\sim0.7$\,dex uncertainty on $M_{\rm BH}$, the small dynamic range for $kT_{\rm disk}$, and unconstrained accretion rate and BH spin effects. In general, we note that both the disk and QPE temperature span a dynamic range of a factor $\approx 2-3$. Thus, it is perhaps unsurprising that they do not show any significant correlation with other quantities.

\section{Summary}

QPEs represent the newest frontier of variable emission from nuclear black holes, and their possible association with low-frequency GW emitters is tantalizing. However, current disk-collision models with either stars or BHs as orbiters suffer from a number of problems and no configuration can currently explain all observed QPE sources \citep[e.g.,][for some recent discussion]{Guo+2025:coll,Mummery+2025:collisions,Yao+2025:simul,Guolo+2025:gsnlongterm}. While current short- and long-term monitoring campaigns are important to constrain this proposed interpretation, discovering new QPE sources is crucial to probe the diversity of the QPE population. 

Here, we report the discovery of a new galaxy showing X-ray quasi-periodic eruptions (Fig.~\ref{fig:lc}), the fifth QPE discovery obtained through a blind search in \emph{SRG}/eROSITA all-sky survey data. The characteristic energy evolution shown by other confirmed QPE sources is recovered (Fig.~\ref{fig:energy_evolution}), thus we name J0325 as eRO-QPE5. eRO-QPE5 shows X-ray properties similar to other QPE sources. In particular, QPE duration and recurrence time ($\sim0.6$\,d and $\sim3.7\,$d, respectively), and integrated energy ($\sim3.4 \times 10^{47}\,$erg) sit at the high end of the known population. The black hole mass, inferred from galaxy stellar mass scaling relations and SED fitting (Fig.~\ref{fig:sed}), is $M_{\rm BH}=2.9^{+5.4}_{-2.2}\times10^7\,M_{\astrosun}$, thus among the highest values together with eRO-QPE4. In agreement with other eROSITA-discovered QPEs, no photometric optical-IR variability is found in archival ASASSN, ATLAS, and NEOWISE data, no radio counterpart from neither the nucleus nor the galaxy is found at tens of $\mu$Jy level, and the optical spectra look featureless (Fig.~\ref{fig:opt_spec}). With a spec-z of $0.1155$, eRO-QPE5 is the farthest QPE source discovered to date.

Given the number of recent discoveries, enough bona fide QPE sources have data available to test for possible correlations. We do not find any correlation between neither $t_{\rm dur}-M_{\rm BH}$ nor $t_{\rm recur}-M_{\rm BH}$, nor between any timing and spectral quantity and the peak disk temperature or the peak QPE temperature (Sect.~\ref{sec:correlations2}). We argue these tests are fundamentally limited by the short dynamic range spanned by the fitted disk and QPE temperatures, and the high statistical and systematic uncertainties of $M_{\rm BH}$ estimates. We confirm the presence of a tight correlation between $t_{\rm dur}$ and $t_{\rm recur}$ as reported in previous work. The $t_{\rm dur}-t_{\rm recur}$ relation is found at slopes $\approx 1$ ($1.14\pm0.16$ in this work), which is consistent with star-disk collisions if powered by stellar debris liberated by previous collisions \citep{Yao+2025:simul}, but there are a few caveats in interpreting the observed relation and any prediction from simplistic accretion disk scalings (Sect.~\ref{sec:correlations}). Interestingly, recently \citet{Guo+2025:coll} and \citet{Mummery+2025:collisions} argued that the high integrated energy observed in some QPE sources cannot be achieved with the star or BH's cross section, and indeed \citet{Mummery+2025:collisions} suggested that it is only achieved if collisions are powered by stellar debris. In turn, an interpretation invoking stellar streams struggles to explain the short period regular QPE sources, such as GSN\,069 and eRO-QPE2 \citep[e.g., see the discussion in][]{Yao+2025:simul}. Thus, currently no single model setup is able to reproduce all QPE sources and observables, perhaps suggesting that there may be a variety of orbiters and orbital configurations in place, or a missing key ingredient to the theory all-together. Further analysis and dedicated simulations are required to shed light on this problem. 

Finally, we note that with eROSITA's operations being currently halted, there is no strong candidate to take over in the systematic and blind discovery of QPEs. Since most known QPE sources peaked in the $\sim (\rm few \times 10^{-13} - \rm few \times 10^{-12})$\,erg\,s$^{-1}$\,cm$^{-2}$ range and their volumetric rates are relatively low \citep{Arcodia+2024:rates}, large search volumes are required with instantaneous depth preferred over large instantaneous area. This will be achievable with AXIS \citep{Reynolds+2023:axis} and NewAthena \citep{Nandra+2013:athena,Cruise+2025:newathena} in the years to come. In the meantime, systematic X-ray follow-up of TDEs and optical transients from MBHs may still provide more QPE sources for further study of this exotic population. For instance, \citet{Chakraborty+2025:upj} recently estimated that of order $\approx 9\%$ of optically-selected TDEs should also, eventually, be X-ray QPE emitters.

\begin{acknowledgments}
We thank the referee for their constructive comments which improved the manuscript. R.A. is grateful to Itai Linial, Giovanni Miniutti, Andrew Mummery, and Mariusz Gromadzki for useful discussions, and to Norbert Schartel and the \emph{XMM-Newton} team for prompt scheduling of the observation. R.A. was supported by NASA through the NASA Hubble Fellowship grant \#HST-HF2-51499.001-A awarded by the Space Telescope Science Institute, which is operated by the Association of Universities for Research in Astronomy, Incorporated, under NASA contract NAS5-26555. A.J.G. is grateful for support from the Forrest Research Foundation. G.P. acknowledges financial support from the European Research Council (ERC) under the European Union’s Horizon 2020 research and innovation program "Hot Milk" (grant agreement No. 865637) and support from Bando per il Finanziamento della Ricerca Fondamentale 2022 dell’Istituto Nazionale di Astrofisica (INAF): GO Large program and from the Framework per l’Attrazione e il Rafforzamento delle Eccellenze (FARE) per la ricerca in Italia (R20L5S39T9). M.K. acknowledges support from DLR grant FKZ 50 OR 2307.

This work is based on data from eROSITA, the soft X-ray instrument aboard SRG, a joint Russian-German science mission supported by the Russian Space Agency (Roskosmos), in the interests of the Russian Academy of Sciences represented by its Space Research Institute (IKI), and the Deutsches Zentrum für Luft- und Raumfahrt (DLR). The SRG spacecraft was built by Lavochkin Association (NPOL) and its subcontractors, and is operated by NPOL with support from the Max Planck Institute for Extraterrestrial Physics (MPE). The development and construction of the eROSITA X-ray instrument was led by MPE, with contributions from the Dr. Karl Remeis Observatory Bamberg \& ECAP (FAU Erlangen-Nuernberg), the University of Hamburg Observatory, the Leibniz Institute for Astrophysics Potsdam (AIP), and the Institute for Astronomy and Astrophysics of the University of Tübingen, with the support of DLR and the Max Planck Society. The Argelander Institute for Astronomy of the University of Bonn and the Ludwig Maximilians Universität Munich also participated in the science preparation for eROSITA. The eROSITA data shown here were processed using the eSASS software system developed by the German eROSITA consortium. Some of the observations reported in this paper were obtained with the Southern African Large Telescope (SALT) under program 2021-2-LSP-001 (PI: D.\ Buckley). Polish participation in SALT is funded by grant No.\ MEiN nr 2021/WK/01. This work made use of data supplied by the UK Swift Science Data Centre at the University of Leicester. 
The Australia Telescope Compact Array is part of the Australia Telescope National Facility (grid.421683.a) which is funded by the Australian Government for operation as a National Facility managed by CSIRO. We acknowledge the Gomeroi people as the traditional owners of the Observatory site.

\end{acknowledgments}

%



\software{astropy \citep{2018AJ....156..123A}, 
          eSASS \citep{Brunner+2022:esass},
          swifttools.ukssdc for light curves \citep{Evans+2007:xrtlc,Evans+2009:xrtlc},
          }



\appendix

\section{Details on data processing and analysis}
\label{sec:xray_proc}

The processing and analysis of data follows closely the one presented in \citetalias{Arcodia+2024:eroqpes} for eRO-QPE3 and eRO-QPE4. Here, we only highlight more details unique to the analysis of eRO-QPE5 data. We adopt a Galactic column density of $1.37\times10^{20}\,$cm$^{-2}$ from \citet{Willingale+2013:nh}. We model the eruption component with a simple black body, which provides a good approximation for the exponential decay shape that is within the observed energy band, despite radiation transport calculations of disk collisions models predict a shallower slope at low-energies \citep{Vurm+2024:emission}.

\subsection{eROSITA}
\label{sec:ero_proc}

Here, we expand on the possible X-ray variability shown by J0325 during eRASS1-3. J0325 was detected in the cumulative image of the first three surveys (eRASS:3), with a flux $F_{0.2-2.0\,\rm keV} = 6.6^{+2.6}_{-1.9}\times 10^{-14}\,$erg\,s$^{-1}$\,cm$^{-2}$ (\texttt{diskbb} model), which prompted deeper investigation. Individual light curves were first rebinned with a standard cut at fractional exposures $f_{\rm exp} >10\%$. $f_{\rm exp}$ is the product of the fractional nominal on-axis effective collecting area and the fraction of the good time interval for which the source was visible in the field of view \citep[see more details in][]{Brunner+2022:esass}. Effectively, $f_{\rm exp}$ is used to convert counts to count rates by multiplying the time bin at the denominator. Lower $f_{\rm exp}$ are found at the start and end of light curves when the source is only partially covered by the edge field of view, which produces a characteristic `smiley' feature in constant light curves. Essentially, rates and rates uncertainties increase at the edges of the light curves, for a constant number of counts, if $f_{\rm exp}$ is small. We refer to \citet{Bogensberger+2024:variab} for more discussion and simulations. 

\begin{figure}[tb]
		\centering
		\includegraphics[width=0.99\columnwidth]{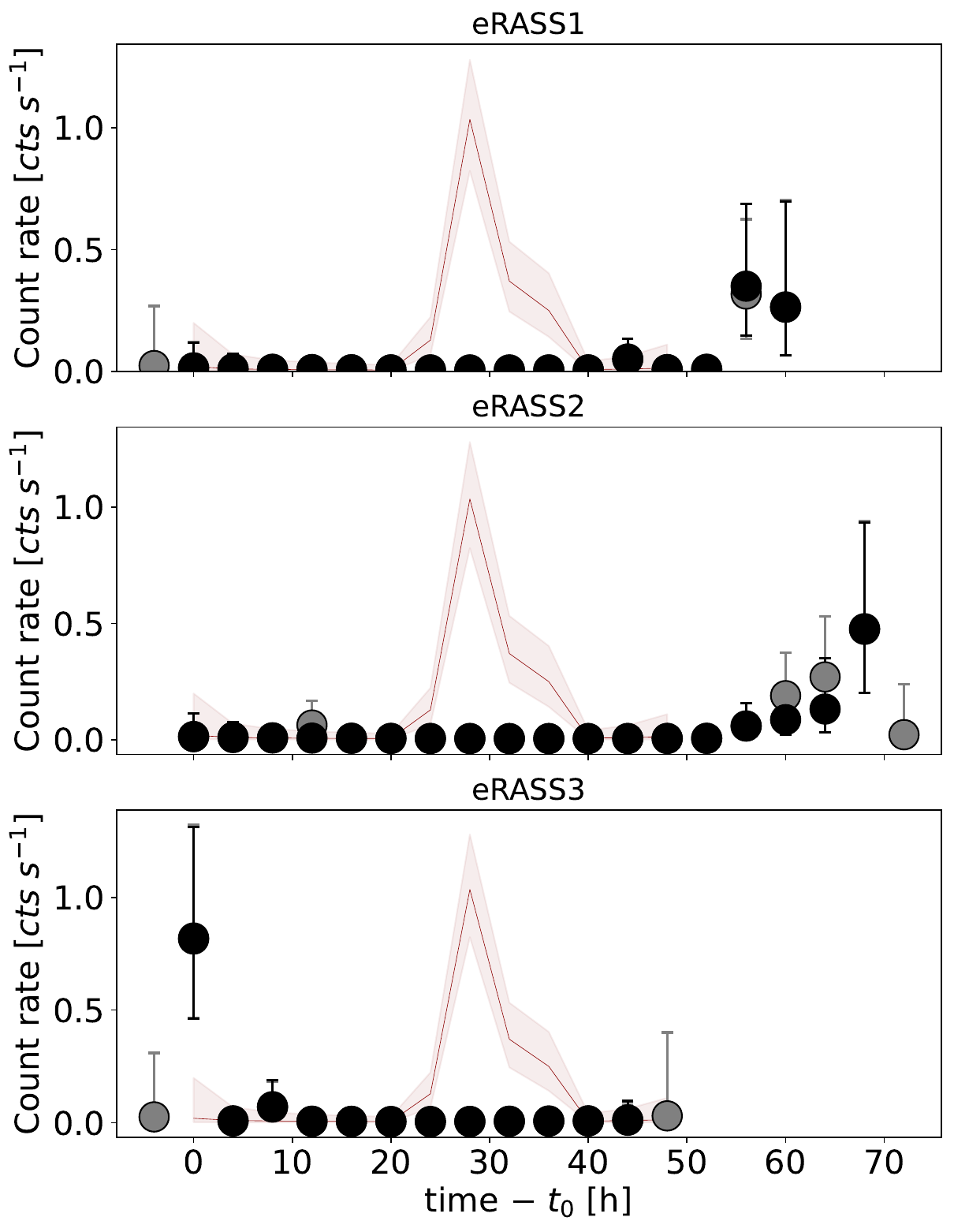}
		\caption{eROSITA $0.2-2.3$\,keV light curves of eRASS1, eRASS2 and eRASS3 (from top to bottom). In black the rebinned light curve including all visits with fractional exposure $f_{\rm exp}\geq 0.1$ and in gray $f_{\rm exp}\geq 0.03$. For comparison, we show in dark red the silhouette of the eRASS4 eruption.}
		\label{fig:erass123}
\end{figure}

Individual eRASS1-2-3 light curves are shown in Fig.~\ref{fig:erass123}, from top to bottom, respectively. The default selection cut at $f_{\rm exp} \geq 10\%$ is shown in black, which shows some counts at the end of eRASS1 and eRASS2, and in one visit at the start of eRASS3. This motivated further investigation of the edges of the individual light curves, and we produced a light curve including all visits with $f_{\rm exp} \geq 3\%$ (shown in gray in Fig.~\ref{fig:erass123}). The eRASS1 light curve remains unaffected, while the last visit of eRASS2 and the first of eRASS3 may suggest that the source was detected in a transient fashion also in these epochs. However, as mentioned above the first and last visits at lower fractional exposures require more in-depth analysis. We thus performed aperture photometry in the good time intervals when the source shows non-zero counts in the eRASS1, eRASS2 and eRASS3 light curves. We computed the binomial no-source probability $P_b$ with the code presented in \citet{Arcodia+2024:ap_photom} and obtained $P_b\sim2.2\times 10^{-9}$, $\sim2.5\times 10^{-12}$, $\sim3.5\times 10^{-9}$ that the detected counts represent a background fluctuation, for eRASS1, eRASS2, and eRASS3, respectively. Thus, the source appears to be significantly detected during those visits. To investigate whether these transients detections are consistent with fluctuations from a constant point source (mimicking the no-QPE quiescence phase), we performed light curve simulations making use of the \texttt{SIXTE}\footnote{\url{https://github.com/thdauser/sixte}} software package (\cite{dauser2019sixte}) provided by ECAP/Remeis observatory. We simulated mock light curves assuming a constant source with $F_{0.2-2.3\,\rm keV} = 1 \times10^{-13}$\,erg\,s$^{-1}$\,cm$^{-2}$ (roughly matching the possible quiescence based on the eRASS non detections and the subsequent \emph{XMM-Newton} detection), using a source spectrum described by a \textsc{diskbb} model and absorbed by galactic N$_\mathrm{H}$. We used the \emph{SRG/eROSITA} attitude files to generate 500 light curves per eRASS and processed them with the same prescriptions as that used on the real data. This allowed us to test the hypothesis that a constant quiescent emission can give rise to patterns similar to those observed, shown in Fig.~\ref{fig:erass123}. The majority of visits show a measured count rate of zero c/s, but 90\% of light curves show at least one day with a count rate above 0.1\,c/s. In the case of eRASS1 and eRASS2, we looked for occurrences of consecutive visits being above the observed threshold, to match the observed light curve (top and middle panel of Fig.~\ref{fig:erass123}). For instance, the eRASS1 light curve shows two consecutive visits with $\sim0.35$\,c/s, $\sim0.27$\,c/s, while eRASS2 shows four consecutive visits with count rate of $\sim0.06$\,c/s, $\sim0.09$\,c/s, $\sim0.13$\,c/s, and $\sim0.48$\,c/s, respectively. From the simulated light curves we find that the probability of having three consecutive visits with a measured count rate around or above $\sim 0.1$\,c/s is 3\%, and similarly, the probability of having two consecutive visits around and above $\sim0.15$\,c/s is again 3\%. In the case of eRASS3, we look for single visits above an observed threshold similar to the observed count rate of $\sim0.8$\,c/s (bottom panel of Fig.~\ref{fig:erass123}). We find that no simulated light curve shows an eROday with a count rate above $\sim0.6$\,c/s, although in 6\% of the light curves there is at least one visit with a count rate above $\sim0.5$\,c/s, which is compatible within uncertainties with the observed count rate. Thus, while it is possible to produce observed patterns like those in Fig.~\ref{fig:erass123} from a constant source at $F_{0.2-2.3\,\rm keV} = 1 \times10^{-13}$\,erg\,s$^{-1}$\,cm$^{-2}$, it is highly unlikely, particularly so considering a transient detection in each eRASS. However, while both our light curves simulations and aperture photometry confirm the likelihood of transient source detection in eRASS1, eRASS2 and eRASS3, we are not able to distinguish between the cases of a variable quiescent emission (with no QPEs), and a constant source in quiescence plus QPEs. In the latter case, the difference compared to the better-resolved eRASS4 eruption can be attributed to vignetting at the edges of the eRASS1-3 light curves. 

\begin{table*}[tb]
	\small
    \centering
	\setlength{\tabcolsep}{5pt}
	\renewcommand{\arraystretch}{1.1}
	\caption{Spectral fit results for eRO-QPE5.}
	\label{tab:spec}
	\begin{threeparttable}
		\begin{tabular}{cccccccc}
			\toprule
            \multicolumn{1}{c}{Epoch} &
			\multicolumn{1}{c}{Spectrum} &
			\multicolumn{1}{c}{Model} &
			\multicolumn{1}{c}{$kT_{\rm disk}$} &
			\multicolumn{1}{c}{$F^{\rm disk}_{\rm 0.2-2.0\,keV}$} &
			\multicolumn{1}{c}{$kT_{\rm QPE}$} &
			\multicolumn{1}{c}{$F^{\rm QPE}_{\rm 0.2-2.0\,keV}$} &
            \multicolumn{1}{c}{$F^{\rm QPE}_{\rm bol}$}
            \\
              &       &          & [eV]   & [erg\,s$^{-1}$\,cm$^{-2}$]  &  [eV]   &  [erg\,s$^{-1}$\,cm$^{-2}$] & [erg\,s$^{-1}$\,cm$^{-2}$] 
            \\
            \midrule
             eRASS4 & Quiesc.      &     \texttt{disk}       &     --   &  $<4.0\times 10^{-15}$   &  --  &  --  & -- \\
              & QPE      &     \texttt{bb}       &     --   &  --   &  $111_{-11}^{+14}$   &  $1.3_{-0.3}^{+0.3} \times 10^{-12}$ & $1.5_{-0.3}^{+0.5} \times 10^{-12}$ \\
              \midrule
                NICER & QPE rise    &     \texttt{bb}       &     --  &  --      &  $(93\pm4)$   &  $3.4_{-0.2}^{+0.2} \times 10^{-13}$  & $4.0_{-0.3}^{+0.3} \times 10^{-13}$ \\
               & QPE peak    &     \texttt{bb}       &     --  &  --    &  $(108\pm3)$   &  $7.0_{-0.3}^{+0.3} \times 10^{-13}$  & $7.9_{-0.3}^{+0.3} \times 10^{-13}$ \\
               & QPE decay    &     \texttt{bb}       &     --   &  --   &  $88_{-0.4}^{+0.3}$   &  $4.5_{-0.2}^{+0.3} \times 10^{-13}$  & $5.5_{-0.3}^{+0.4} \times 10^{-13}$ \\
              \midrule
              XMM & Quiesc.      &     \texttt{disk}       &     $37^{+11}_{-7}$   &  $5.4^{+2.3}_{-1.8}\times 10^{-13}$    &  --  & -- & --\\
                  & QPE rise1    &     \texttt{disk+bb}       &     $\sim37$   &  $\sim5.4\times 10^{-14}$   &  $(78\pm4)$   &  $2.7_{-0.2}^{+0.3} \times 10^{-13}$  & $3.8_{-0.5}^{+0.6} \times 10^{-13}$\\
                 & QPE rise2    &     \texttt{disk+bb}       &     $\sim37$  &  $\sim5.4\times 10^{-14}$      &  $(99\pm2)$   &  $5.5_{-0.1}^{+0.2} \times 10^{-13}$  & $6.8_{-0.2}^{+0.2} \times 10^{-13}$ \\
               & QPE peak    &     \texttt{disk+bb}       &     $\sim37$  &  $\sim5.4\times 10^{-14}$    &  $109_{-1}^{+2}$   &  $8.7_{-0.2}^{+0.1} \times 10^{-13}$  & $1.02_{-0.02}^{+0.03} \times 10^{-12}$ \\
               & QPE decay1    &     \texttt{disk+bb}       &     $\sim37$   &   $\sim5.4\times 10^{-14}$   &  $92_{-2}^{+1}$   &  $7.3_{-0.3}^{+0.2} \times 10^{-13}$  & $9.3_{-0.4}^{+0.3} \times 10^{-13}$ \\
               & QPE decay2    &     \texttt{disk+bb}       &     $\sim37$   &   $\sim5.4\times 10^{-14}$    &  $57_{-3}^{+1}$   &  $2.3_{-0.1}^{+0.2} \times 10^{-13}$   & $4.7_{-0.5}^{+0.6} \times 10^{-13}$ \\
             \bottomrule
		\end{tabular}
	    \begin{tablenotes}
        \small
        \item Fit values show the median and related $16^{\rm th}-84^{\rm th}$ percentiles of the fit posteriors, or the $84^{\rm th}$ percentile only for upper limits. Fluxes are shown unabsorbed and in the rest frame. Here we show only spectral fits on X-ray data alone (e.g., continuum models are \texttt{zashift * diskbb} for the quiescent disk and \texttt{zbbody} for the QPEs). During the QPE fit, the quiescence model component is held fixed at the quiescence-only fit given the low signal-to-noise ratio. For \emph{XMM-Newton}, the different phases are shown in Fig.~\ref{fig:energy_evolution}. Given the spectroscopic redshift adopted ($z=0.1155$) and the cosmology adopted \citep{Hinshaw+2013:wmap9}, the conversion for related luminosity values for eRO-QPE5 is $3.53\times10^{55}$\,cm$^2$ in this paper.
        \end{tablenotes}
   \end{threeparttable}
\end{table*}

We conclude that it is possible that the source was emitting eruptions at the eRASS1, eRASS2, and eRASS3 epochs. This would increase the QPE lifetime by $\sim 1.5\,$y. 
Under this assumption, we isolated the visits with a detection from those compatible with background (using the more conservative light curve with $f_{\rm exp}\geq 0.1$), and show the related phases in the long-term evolution plot accordingly (Fig.~\ref{fig:longterm}). For completeness, we show in gray the fluxes inferred from the total exposure without separating bright and quiescence phases (a detection and a non-detection, respectively).

%

\subsection{NICER}

We followed the time-resolved spectroscopy approach for reliable estimation of source light curves outlined in Section 2.1 of \cite{Chakraborty+2024:ero1}. To summarize, we split data into observations of a few hundred seconds---the characteristic timescale for significant background variability---and perform independent, joint spectroscopic modeling of the source and background each window. Spectral fitting and background estimation was performed with the \texttt{SCORPEON}\footnote{\href{https://heasarc.gsfc.nasa.gov/docs/nicer/analysis_threads/scorpeon-overview/}{https://heasarc.gsfc.nasa.gov/docs/nicer/\\analysis\_threads/scorpeon-overview}} model over a broadband energy range (0.25--10 keV) for data taken in orbit night, and a slightly restricted range (0.38--10 keV) during orbit day. \texttt{SCORPEON} is a semi-empirical, physically motivated background model which explicitly includes components for the cosmic X-ray background as well as non X-ray noise events (e.g. precipitating electrons and cosmic rays) and can be fit along with the source to allow joint estimation of uncertainties. We grouped our spectra with the optimal binning scheme of \cite{Kaastra16}, i.e. \texttt{grouptype=optmin} with \texttt{groupscale=10} in the \texttt{ftgrouppha} command, and performed all spectral fitting with the Cash statistic \citep{Cash1979}.

We fitted the observed eruptions (Fig.~\ref{fig:lc_fit}) using UltraNest \citep{Buchner2019:mlf, Buchner2021:ultranest} with three phenomenological models often used in QPEs: a double exponential for both rise and decay \citep[`model1'; e.g.,][]{Arcodia+2022:ero1_timing}, a Gaussian rise and exponential decay (`model2'), and a Gaussian profile (`model3'), in addition to a constant. We fitted each eruption separately and compared the logarithm of the Bayesian evidence ($\log Z$) to find the best-fit model. All three models yield comparable $\log Z$ values for the first and third eruption, while the double exponential model shows the lowest $\log Z$ value for the second eruption. Based on the evident asymmetry in the deeper \emph{XMM-Newton} observation, and that `model1' reproduces better the peak time bin with its modeled peak, we adopt `model1' as best-fit model. For more details on the shape and computation of rise and decay times we refer to \citet{Arcodia+2022:ero1_timing}. In summary, we computed the burst rise-to-decay duration from the times at which the fitted profile reaches $1/e^3$ of the peak, and the recurrence time as peak-to-peak time separation. Finally, the quiescent emission in between eruptions is not detected by \emph{NICER}. Thus, we extracted spectra during three phases (rise, peak and decay), qualitatively isolating the brightest time bins, and fitted them with a simple black body model for the eruption component \citep[e.g., see][]{Chakraborty+2025:upj}. The characteristic harder rise and softer decay is recovered, even if within larger uncertainties due to the lower signal-to-noise ratio: namely $kT\sim93\,$eV, $kT\sim108\,$eV, and $kT\sim88\,$eV, for rise, peak, and decay, respectively (see Table~\ref{tab:spec}).

\subsection{XMM-Newton}

We reduced \emph{XMM-Newton} data of EPIC-MOS1 and 2 \citep{Turner+2001:mos} and EPIC-PN \citep{Struder+2001:EPIC} cameras and the Optical Monitor \citep[OM;][]{Mason+2001:OM} using standard tools and prescriptions (SAS v. 20.0.0 and HEAsoft v. 6.32). Event files from EPIC cameras were screened for flaring particle background, which results in $\sim52$\,ks of net exposure over a baseline of $\sim100$\,ks. Source (background) regions were extracted within a circle of $43^{\prime\prime}$ centered on the source (in a nearby source-free region). We refined the X-ray position using the task \texttt{eposcorr}, obtaining a 1$\sigma$ positional circle (shown in green in Fig.~\ref{fig:opt_image}) with an accuracy of $2.6^{\prime\prime}$. 

\begin{figure}[tb]
		\centering
		\includegraphics[width=0.99\columnwidth]{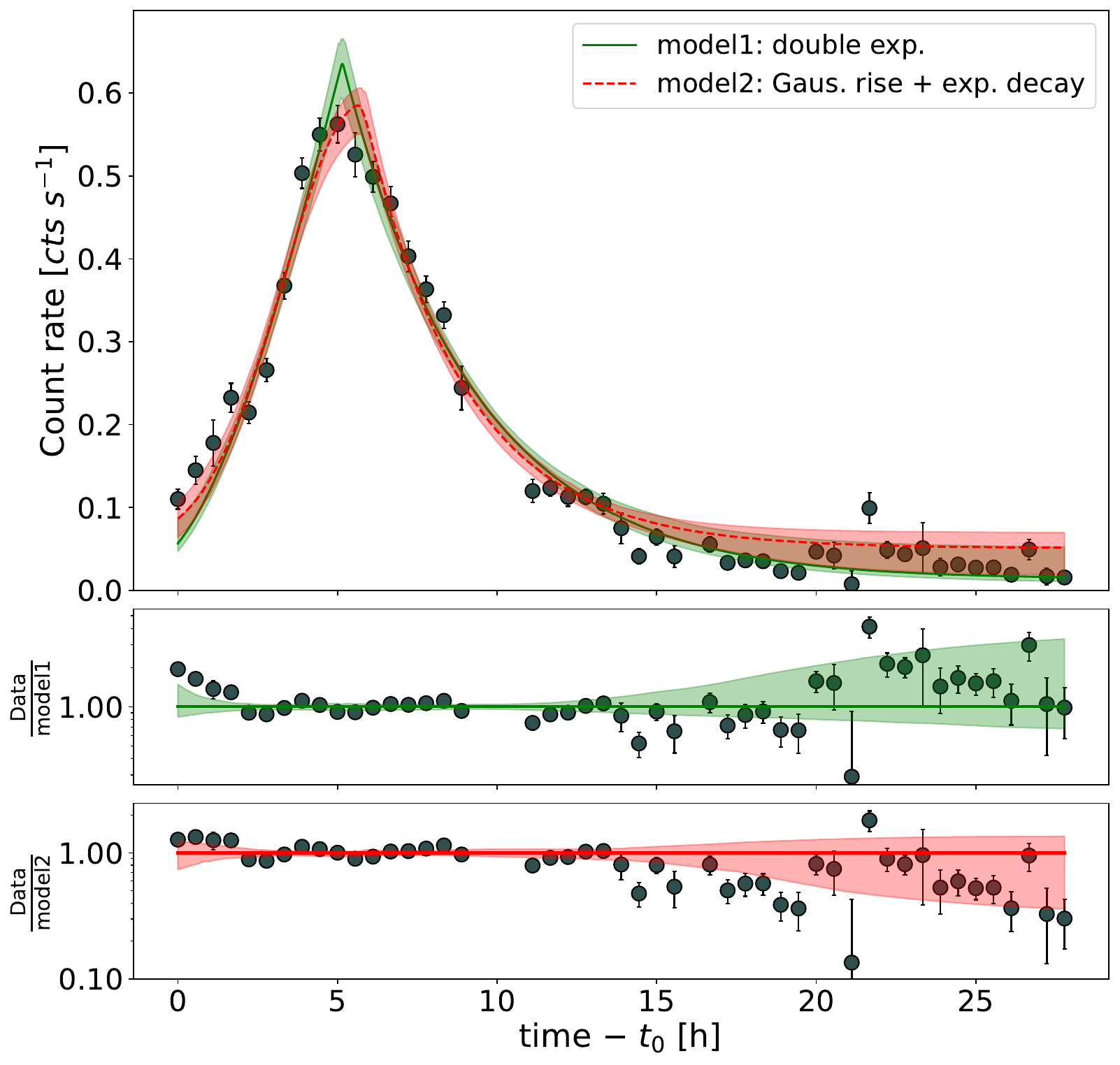}
		\caption{\emph{XMM-Newton} light curve and two models with their residuals, as shown in the legend. The model is shown with a line (median) and related percentile contours (equivalent to $3\sigma$). 
        }
		\label{fig:lc_fit}
\end{figure}

The resulting $0.2-2.0\,$keV EPICpn light curve is shown in the bottom panel of Fig.~\ref{fig:lc}, which contains almost a full eruption and some signal in quiescence. We fitted the light curve profile using the same three models adopted for the \emph{NICER} light curve, in addition to a constant representing the quiescence level, here detected by \emph{XMM-Newton} (see below). None of them reproduces the entire burst profile (which is unusual compared to other QPE sources), although the exponential rise and decay \citep[e.g.,][]{Arcodia+2022:ero1_timing} obtains the best score and overall residuals, with an improvement $\Delta \log Z \sim 4$ compared to `model2' and $\sim13$ compared to `model3'. Even the best-fit model among these, however, does not reproduces the peak flux accurately (Fig.~\ref{fig:lc_fit}). Since similarly bad residuals were found in eRO-QPE1 \emph{XMM-Newton} data, then corrected adding more burst profiles to the fit \citep{Arcodia+2022:ero1_timing}, we followed the same approach. However, fitting the \emph{XMM-Newton} eruption with two or three burst profiles does not improve the fit. Furthermore, given the high background flares in part of the observations, we checked different background flare thresholds, and different energy ranges (e.g., $0.2-1.0\,$keV and $0.2-0.6\,$keV), although the residuals persist and are thus intrinsic to the source. We defer to future work for a detailed investigation on the burst profile in eRO-QPE5 for high-resolution \emph{XMM-Newton} data. For now, we adopt `model1' and note that the peak time seems to be reproduced relatively well (the peak time from `model1' is the closest to the naif estimate inferred from the brightest time bin). This supports further the use of `model1' for \emph{NICER} data.

\subsection{UV data and spectral energy distribution}
\label{sec:sed}

Here, we provide more details on the analysis of optical-UV data and SED fitting. \emph{Swift}/UVOT data were taken in ``Filter of the Day'' mode throughout the XRT baseline, namely with the UVW2 (208.4\,nm), UVM2 (224.5\,nm), and UVW1 (268.2\,nm) filters, respectively. A light curve was produced with the task \texttt{uvotmaghist} using a standard 5\,arcsec aperture (Fig.~\ref{fig:om}). The related mean standard deviation magnitude values are $19.25\pm0.22$, $19.11\pm0.17$, and $19.13\pm0.20$, for the UVW1, UVM2, UVW2 filter, respectively. These values correspond to fluxes of $\lambda F_{\rm \lambda} = (2.2 \pm 0.4)\times 10^{-13}\,$erg\,s$^{-1}$\,cm$^{-2}$, $(2.4 \pm 0.4)\times 10^{-13}\,$erg\,s$^{-1}$\,cm$^{-2}$, and $(2.5 \pm 0.5)\times 10^{-13}\,$erg\,s$^{-1}$\,cm$^{-2}$, respectively. In addition, several $\sim4400$\,s exposures were taken with the UVW1 filter (291\,nm) of the Optical Monitor \citep[OM;][]{Mason+2001:OM} bracketing the EPIC exposures. While the initial out-of-the-box light curves from both the imaging and timing data (outputs of standard \texttt{omichain} and \texttt{omfchain}) showed some possible variability, more careful analysis finds this source to be constant throughout. In the timing data (\texttt{omfchain}), we attributed the spurious variability to the source being somewhat extended compared to the small fast-mode photometry window and only partially covered by it. In the imaging data, the apparent variations in the computed photometry with the standard \texttt{omichain} command (show in grey in Fig.~\ref{fig:om}, top panel) are attributed to the different source size inferred across exposures, mainly due to the noisy scattered light environment near the center of the field of view. Thus, we recomputed the source magnitudes with the task \texttt{omphotom} choosing aperture photometry as extraction mode and using the same size aperture for each image (5.7\,arcsec). In addition, exposures S011 and S013 are affected by data-loss from the image close to the target (also driving spurious variability) and are thus excluded from our analysis. We show the reliable constant source photometry in Fig.~\ref{fig:om}, with a mean (and standard deviation) magnitude of $18.97\pm0.10$. We converted this to a mean 291\,nm flux (and standard deviation) of $\lambda F_{\rm \lambda} = (2.8 \pm 0.3)\times 10^{-13}\,$erg\,s$^{-1}$\,cm$^{-2}$.

\begin{figure}[t]
		\centering
		\includegraphics[trim={0 0 1.5cm 1.4cm},clip,width=0.99\columnwidth]{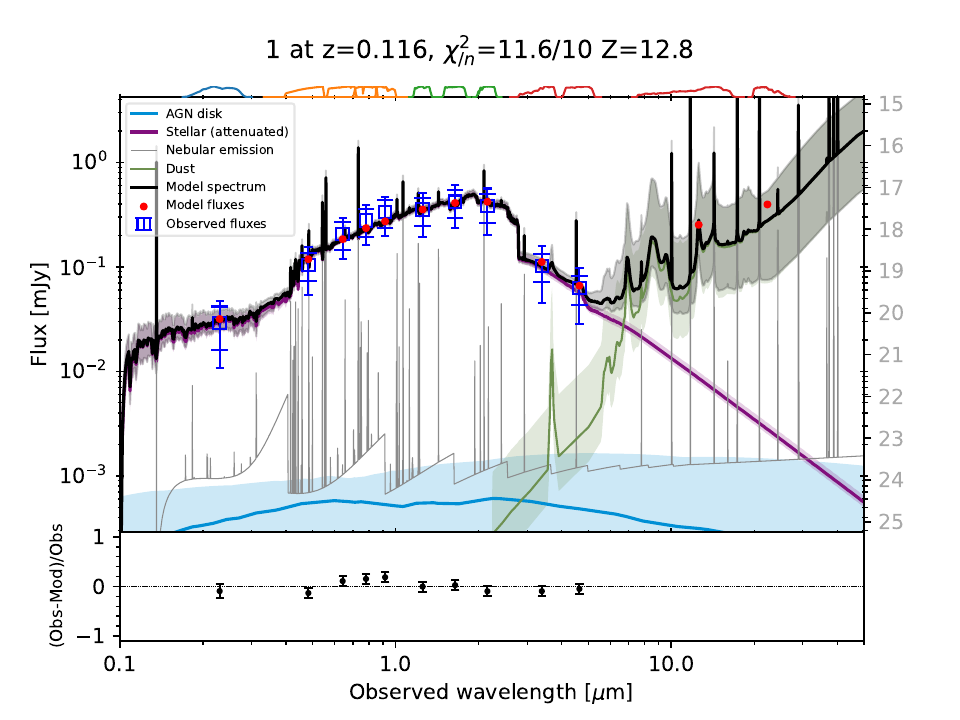}
		\caption{Same as Fig.~\ref{fig:sed}, but in mJy units.}
		\label{fig:sedmJy}
\end{figure}

For the SED fitting with GRAHSP \citep{Buchner+2024:GRAHSP} shown in Fig.~\ref{fig:sed} and Fig.~\ref{fig:sedmJy}, we adopted a prior of $\log (\lambda L_{5100\AA}) = 41$ with a lower uncertainty of 0.5\,dex and an upper uncertainty of 0.3\,dex. This AGN disk level roughly matches that of the observed UV disk emission in eRO-QPE2 \citep[e.g.,][]{Wevers+2025:sed} and it is also within the typical range of $L_{\rm UV}$ in TDE accretion disks (A. Mummery, priv. comm.). It is also compatible with the mean value fitted from the run with a uniform prior on the $L_{5100\AA}$, but in this case the posterior spans $\log \lambda L_{5100\AA} \sim 38-42$. As we know from the detected soft X-rays in quiescence that a disk is present, its luminosity cannot be much lower than $\log \lambda L_{5100\AA} \sim 40$ based on naive extrapolations, thus the posterior solutions at very low AGN disk luminosities have the effect of increasing the upper range of the $M_*$ posterior. This is a small effect, as the mean $M_*$ with uniform $L_{5100\AA}$ is $M_* = 8.8^{+ 2.5}_{- 1.2} \times 10^{9} M_{\astrosun}$, compared to $M_* \sim 8.5\times 10^{9} M_{\astrosun}$ using the $L_{5100\AA}$ prior. Furthermore, we tested our prior by broadening the uncertainties to a lower 1\,dex and an upper 0.5\,dex, and we find small differences within the uncertainties of the fit ($M_* \sim 8.1\times 10^{9} M_{\astrosun}$).

\section{Optical spectra}
\label{sec:app_opt}

We took a spectrum with the Robert Stobie Spectrograph (RSS, \citealp{Burgh+2003:RSS}) on the Southern African Large Telescope (SALT, \citealp{Buckley+2006:SALT}) on the night of 24 February 2023 (Fig.~\ref{fig:opt_spec}). The PG900 VPH grating was used to obtain two 300\,s exposures at different grating angles, allowing for a total wavelength coverage of $\sim3500-7400\AA$. Data were reduced following the procedure outlined in \citetalias{Arcodia+2021:eroqpes} and \citetalias{Arcodia+2024:eroqpes}. The spectrum appears overall featureless, with a tentative identification of a redshift of $z=0.1155$ based on the possible alignment of Calcium absorption line, [OII] in emission and G-band absorption (as shown in the zoom-in). Despite the low signal-to-noise ratio, this value is in the ball park of photometric redshifts in the literature \citep[ranging $\sim0.12-0.14$;][]{Dalya+2022:GLADE+,Duncan2022:ls8_photoz}. Thus, we adopt $z=0.1155$ as distance estimate for eRO-QPE5. While detailed line diagnostics are not possible, this would exclude the presence of a pre-existing strong AGN in the nucleus, in agreement with other QPE sources.

\begin{figure*}[t]
		\centering
        \includegraphics[width=0.99\textwidth]{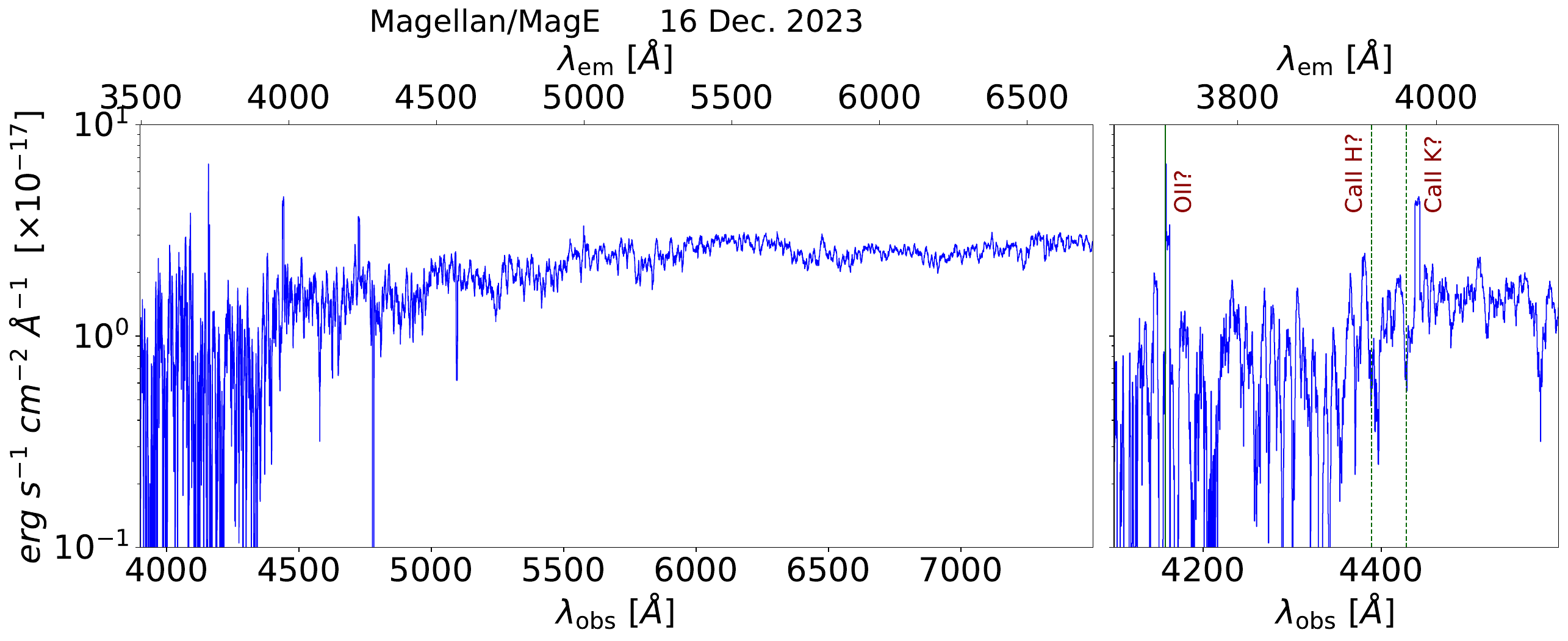}
    	\caption{Magellan/MagE spectrum, which appear noisy and featureless. The zoom-in shows the same region with the expected location of absorption/emission lines based on the SALT redshift estimate (Fig.~\ref{fig:opt_spec}).}
		\label{fig:opt_spec_2}
\end{figure*}

We also obtained a spectrum of eRO-QPE5 with the Magellan Echellette (MagE) spectrograph \citep{marshall_mage} on the 6.5m Magellan/Baade telescope at Las Campanas Observatory on 16 December 2023. The 0.5\arcsec\, slit was used with a total exposure time of 45 minutes. The data were reduced using PypeIt \citep{prochaska_pypeit} with standard reduction procedures, including wavelength calibrations with ThAr lamp exposures and flux calibrations with observations of a standard star taken on the same night. The MagE spectrum is also relatively featureless and noisy, with tentatively similar Calcium absorption lines (Fig.~\ref{fig:opt_spec_2}).

\bibliography{sample631}{}
\bibliographystyle{aasjournal}



\end{document}